\documentclass[a4paper,11pt]{article}
\pdfoutput=1 % if your are submitting a pdflatex (i.e. if you have
\usepackage{jheppub} % for details on the use of the package, please
\usepackage[T1]{fontenc} % if needed
\usepackage{epsfig}
\usepackage{amsmath}
\usepackage{bbm}
\usepackage{amssymb}
\usepackage{amsfonts}
\usepackage{amsbsy}
\usepackage[all]{xy}
\usepackage{mathrsfs}
\usepackage{graphicx}
\usepackage{blindtext}
\usepackage{slashed}
\usepackage{mathtools}
\usepackage{tikz} 
\usepackage{pgfplots}
\usepackage{makecell}
\usepackage{multicol}
\usetikzlibrary{tqft}
\def\be{ \begin{equation} }
\def\ee{ \end{equation}}

\def\Aut{{\rm Aut}}

\def\det{{\rm det}}
\def\dim{{\rm dim}}
\def\End{{\rm End}}
\def\exp{{\rm exp}}

\def\Hom{{\rm Hom}}

\def\log{{\rm log}}

\def\Mat{{\rm Mat}}

\def\Tr{{\rm Tr}}

\def\half{\frac{1}{2}}
\def\p{\partial}

\def\one{{\hbox{ 1\kern-.8mm l}}}
\def\CA {{\cal A}}
\def\CB {{\cal B}}
\def\CC {{\cal C}}

\def\CE {{\cal E}}
\def\CF {{\cal F}}

\def\CH {{\cal H}}

\def\CO {{\cal O}}

\def\CW {{\cal W}}

\def\CO {{\cal O}}
\def\CZ {{\cal Z}}
\def\CE {{\cal E}}

\def\CH {{\cal H}}

\def\CB {{\cal B}}

\def\CT {{\cal T}}

\def\CZ {{\cal Z}}

\def\IC{\mathbb{C}}

\def\IP{\mathbb{P}}

\def\IR{{\mathbb{R}}}

\def\IZ{{\mathbb{Z}}}

\def\fB{\mathfrak{B}}

\def\fd{\mathfrak{d}}

\def\fh{\mathfrak{h}}

\def\fo{\mathfrak{o}}

\def\fr{\mathfrak{r}}

\def\fr{\mathfrak{r}}

\def\fC{\mathfrak{C}}

\def\rmk#1{\bigskip\noindent{\bf Remarks} }
\title{\boldmath Comments On Summing Over Bordisms In TQFT}

\author[a]{Anindya Banerjee and Gregory W. Moore}

\affiliation[a]{NHETC and
$~~$Department of Physics and Astronomy, Rutgers University \\
$~~$126 Frelinghuysen Rd., Piscataway NJ 08855, USA\\
\\}

\emailAdd{ab1702@scarletmail.rutgers.edu}
\emailAdd{gwmoore@physics.rutgers.edu}

\abstract{Recent works in quantum gravity, motivated by the ``factorization problem'' 
and ``baby universes,'' have 
considered sums over bordisms with fixed boundaries in topological
quantum field theory (TQFT). We discuss this construction and observe a
curious splitting formula for the total amplitude. Version two revised: \today }

\begin{document} 
\maketitle
\flushbottom

\section{Statement Of The General Problem}

A long-standing question in quantum theories of
gravity has been the question of whether, and how,
one should sum over topologies of spacetime in the path integral \cite{Banks:1989zw, Banks:1988je, Coleman:1988cy, Fischler:1989ka, Giddings:1988wv}.
Some recent literature in quantum gravity has
considered a modification of topological field theory
wherein the amplitudes, between fixed in and outgoing
boundaries are defined by the sum over all the bordisms
that fill in those boundaries \cite{Balasubramanian:2020jhl,deMelloKoch:2021lqp, Gardiner:2020vjp,Marolf:2020xie}.
While it is a standard idea in physics to sum over topologies,
the sum in the
context of topological field theory has not
been considered before in the mathematical literature on topological field theory, as far as we are aware.
\footnote{Of course, it is absolutely standard in topological string theory to sum
over all bordisms from the emptyset to itself, i.e. the vacuum amplitude. But that
is the trivial part of the present story.}
In this paper we make a few observations about the sum over 
topologies in low-dimensional TQFT's and in particular 
observe an interesting splitting formula for the total amplitude.
\footnote{We use the term ``splitting formula'' and not the more natural term ``factorization formula'' because the latter term has a different meaning in the quantum gravity literature.}

The general problem may be stated as follows: We consider a general $d$-dimensional TQFT in the sense of Atiyah and Segal.
\footnote{This approach to quantum field theory 
is explained in pedagogical terms in many many references. 
In this paper we will be drawing heavily on the description of the two-dimensional 
case in \cite{Aspinwall:2009isa,Moore:2006dw}. We therefore recommend that exposition.}
It is therefore expressed as a functor
\be\label{eq:TQFT}
\CZ: \fB\fo\fr\fd \rightarrow \fC  ~ .
\ee
Here  $\fB\fo\fr\fd = {\rm Bord}_{\langle d,d-1 \rangle}(\CF)$ denotes a monoidal category whose objects are $(d-1)$-dimensional
manifolds and morphisms are $d$-dimensional bordisms. The manifolds are endowed with background fields $\CF$. 
\footnote{We use the term ``background field'' in the sense explained in \cite{Freed:2013gc}.   
Thus a background field could be a bundle with connection, or a tangential structure like an orientation or (s)pin structure. We call a bordism ``connected'' if the underlying manifold is connected.} 
The target category $\fC$ is in general just a symmetric monoidal $d$-category, but
for simplicity we will only consider the ``top three levels'' of the categories. In this paper partition functions are valued in a field $\kappa$ and state spaces associated to a 
$(d-1)$-dimensional manifold are vector spaces over $\kappa$. We will briefly encounter 
the third categorical level in the 2d open-closed case where we use the category $\fB$ of boundary conditions. We will refer to $\CZ$ as the \emph{seed TQFT}. 

We now consider the following question: Fix in-going and out-going objects $X_{\rm in}$ and $X_{\rm out}$ in the bordism
category and consider the sum
\be\label{eq:Sum-TQFT}
\CA(X_{\rm in}, X_{\rm out}) := \sum_{Y: X_{\rm in} \to X_{\rm out}}  \frac{1}{\vert \Aut(Y)\vert} \CZ(Y) ~ .
\ee
which we will refer to as the \emph{summed amplitude}. 
Here we divide the order of the group of automorphisms of the isomorphism
type of $Y$.

Importantly, when dividing by the automorphism group
 we confine this group to automorphisms which are  the \underline{identity on the boundaries}.
This combinatorial rule is extremely important. It is implicitly assumed in
 \cite{Balasubramanian:2020jhl,deMelloKoch:2021lqp,Gardiner:2020vjp,Marolf:2020xie}, so we will work with it. See Figure 1. It might be of interest to relax the constraint of the identity on the boundary, but
 we will not do so in this paper. 
 \footnote{M. Kontsevich has suggested some interesting alternatives.}

\clearpage 
\noindent\rule[0.5ex]{\linewidth}{1pt}
\\
\begin{tikzpicture}[tqft/cobordism/.style={draw}, tqft/every boundary component/.style={draw}]
\pic [tqft/cap, name=a, fill=gray, at={(0,0)}];
\pic [tqft/cup, name=b, fill=gray, at=(a-outgoing boundary 1)];
\pic [tqft/cap, name=c, fill=gray, at={(2,0)}];
\pic [tqft/cup, name=d, fill=gray, at=(c-outgoing boundary 1)];
\pic [tqft/cap, name=e, fill=gray, at={(8,0)}];
\pic [tqft/cap, name=f, fill=gray, at={(10,0)}];
\end{tikzpicture}
\paragraph{} 
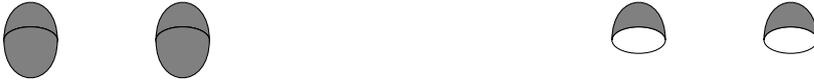
\textbf{Figure 1:} \emph{Here the bordism on the left, which is a bordism from $\emptyset$ to $\emptyset$ and has no boundary,  is weighted with $\frac{1}{2!}$. On the other hand, the surface on the right, 
which has nonempty boundary $S^1 \amalg S^1$, and can be interpreted 
as a bordism in many ways, depending on whether the circles are ingoing or outgoing, is weighted with $1$.} 
\paragraph{} 
\noindent\rule[0.5ex]{\linewidth}{1pt}

The expression in \eqref{eq:Sum-TQFT} is, formally, an element of $ \Hom( \CZ(X_{\rm in} ) , \CZ(X_{\rm out}) ) $
and it is the main object of study in this paper.
A few natural questions about it are:

\begin{enumerate}

\item Does the sum exist?

\item If yes, is it computable and what properties does it have?

\item Is there an extension of the construction to a fully local (a.k.a. fully 
extended) TQFT?

\end{enumerate}

In what follows we will mostly consider $d=1,2$, 
in which case the sum does indeed exist
for a generic theory. We will compute it in many cases, thus recovering and
extending results of   \cite{Balasubramanian:2020jhl,Gardiner:2020vjp,Marolf:2020xie}.

One novel point we observe is a \emph{splitting formula} for the ``total amplitude.'' 
To define the ``total amplitude'' we first remark that the direct sum over all summed
amplitudes is \emph{a priori} defined on the tensor space 
\be
\CA: T^\bullet ( \oplus_X \CZ(X)) \rightarrow T^\bullet ( \oplus_X \CZ(X) ) 
\ee
where, for any vector space $V$ we denote the tensor algebra by $T^\bullet V$ and 
where $\CZ(X)$ is the vector space assigned to (the isomorphism class of)
$X$ and the sum is over all connected objects in $\fB\fo\fr\fd$. (Recall ``connected'' means the underlying manifold is connected.)   Because 
of the sum over all bordisms the amplitudes are totally symmetric and hence 
descend to define an amplitude $\bar\CA$ on the symmetric algebra 
(i.e. they descend to a Fock space) to give a commutative diagram: 
\be\label{eq:SymmetricDescent}
{ \xymatrix{ T^\bullet ( \oplus_X \CZ_X) \ar[r]^{\CA} \ar[d] & T^\bullet ( \oplus_X \CZ_X)\ar[d]\\
 S^\bullet(\oplus_X\CZ_X) \ar[r]^{\bar\CA} &   S^\bullet(\oplus_X\CZ_X) \\ } 
}
\ee
We denote the symmetric algebra of a vector space $V$ by $S^\bullet V$. 
We will mostly focus on obtaining formulae for  
\be
\bar\CA \in \Hom(\CH_{\rm tot}, \CH_{\rm tot} )
\ee
where
\be\label{eq:GenStateSpace}
\CH_{\rm tot} = \otimes_{X} S^\bullet \CZ(X)
\ee
and the tensor product is over (the isomorphism class of) 
all \underline{connected} objects in $\fB\fo\fr\fd$. 
\footnote{There is a theorem \cite{CheegerKister,Wunsch} that there is a countable number of 
diffeomorphism classes of closed connected manifolds of a fixed dimension so we have a countable product of Fock spaces of finite-dimensional Hilbert spaces in equation \eqref{eq:GenStateSpace}. Therefore the domain and codomain of $\bar \CA$ can be defined in a mathematically rigorous way.} 
We will call $\CH_{\rm tot}$ the \emph{Fock space of the seed TQFT $\CZ$.}

In this paper we will show that in the cases $d=1,2$ and $\kappa=\IC$ there
exists an inner product space $\CW$, which we refer to as the 
\emph{splitting space}, such that there exists a map:  
\be
\Phi: \CH_{\rm tot} \to \CW
\ee
with the property that 
\footnote{Given vector spaces $V_1. V_2, V_3$ over a field $\kappa$ there
are two ways to write down the canonical pairing
\be\label{eq:CompMorph}
\Hom(V_1, V_2) \times \Hom(V_2, V_3) \to \Hom(V_1, V_3).
\ee
If $T_{12} \in \Hom(V_1,V_2)$ and $T_{23}\in \Hom(V_2,V_3)$ then 
the pairing is often denoted by the composition of maps:   $T_{23}\circ T_{12}$. However in this paper
we will find it more convenient to use the isomorphism $\Hom(V_1,V_2) \cong V_1^\vee \otimes V_2$ 
and then multiply the operators left to right and contract the inner vector spaces, 
so we will write the pairing \eqref{eq:CompMorph} as $T_{12}T_{23}$ 
which means we take $T_{12}\otimes T_{23} \in V_1^\vee \otimes V_2 \otimes V_2^\vee \otimes V_3$ 
and then use the contraction $V_2\otimes V_2^\vee \to \kappa$ to map to $V_1^\vee \otimes V_3$. }  
\be
\bar\CA = \Phi \Phi^\dagger
\ee
where the $\dagger$ makes use of the sesquilinear inner product on $\CW$
and a natural inner product on $\CH_{\rm tot}$ 
provided by the data of $\CZ$. 
We will see that the inner products are nondegenerate, but the existence of a splitting in general does not require them to be positive definite. 

In fact, the TQFT's $\CZ$ we consider come in continuous families. 
Denoting the data used to define the seed theory $\CZ$ by $\lambda$ 
the total amplitude is a function of $\lambda$ so we write $\bar\CA_{\lambda}$
and similarly 
the map $\Phi$ depends on $\lambda$ so we denote it as 
$\Phi_{\lambda}$. It turns out that $\lambda$ must be ``real,'' 
in a sense described below in examples, so the actual splitting formula is 
\be\label{eq:SplittingFormula}
\bar\CA_{\lambda} = \Phi_{\lambda}\Phi_{ \lambda}^\dagger ~ . 
\ee

A few comments on the splitting formula \eqref{eq:SplittingFormula} are in 
order. First, it is clear that for a fixed $\CW$ the map $\Phi_\lambda$ is not
unique as it can always be right-multiplied by a unitary operator on $\CW$. 
Moreover, the space $\CW$ is itself not unique. 
\footnote{As a simple example of non-uniqueness replace $\CW$ by 
the direct sum of $n$ copies $\CW \oplus \cdots \oplus \CW$
and replace $\Phi$ by $\sqrt{p_1} \Phi \oplus \cdots \oplus 
\sqrt{p_n} \Phi$ where $p_i>0$ and $\sum_i p_i =1$. A natural way to formulate 
the notion of a \emph{minimal} splitting would be to use a universal property. 
That is, to show that there exists a splitting $\Phi_m:\CH_{\rm tot}\to \CW_m$ such that, 
given any other splitting $\Phi: \CH_{\rm tot} \to \CW$  there is an isometry 
$\psi: \CW_m \to \CW $ so that we have a commutative diagram: 
\be\label{eq:minimal-splitting} 
\xymatrix{   & \CW \ar[rd]^{\Phi^\dagger} &  \\ 
\CH_{\rm tot} \ar[ru]^\Phi \ar[rd]_{\Phi_m}  &   &  \CH_{\rm tot} \\
 & \CW_m \ar[uu]^\psi \ar[ru]_{\Phi_m^\dagger} \\ }
\ee
We do not know if such a minimal splitting exists. 
}
The situation is very similar 
to that of the Stinespring representation 
of positive maps of $C^*$ algebras, 
and our splitting formula bears some resemblance to that result. For more precise
versions of \eqref{eq:SplittingFormula}  see  equation \eqref{eq:1d-splittingformula} below
for the $1d$ case, equations \eqref{eq:2d-SS-Splitting-1}
to \eqref{eq:2d-SS-Splitting-3} below for the 2d case with semisimple
closed string Frobenius alebra, and equations \eqref{eq:Open-Amp-Split} and \eqref{eq:Split-Amp-OpenClosed}  below for the case of
semisimple open-closed 2d TQFT.

Since the total amplitude can be expressed 
as an operator-valued inner product it is natural to 
suspect that there is a  role for Hilbert C* modules. 
Thus, the splitting formula will perhaps make contact with some ideas in \cite{Moore:2017wlq}. In the interpretation of \cite{Moore:2017wlq} 
the operator algebra on the Fock space $\CH_{\rm tot}$ described above 
would describe a noncommutative family of quantum systems. The quantum system, whose ``Hilbert space'' is the splitting space $\CW$ is,  
in a (debatable) sense, ``holographically dual'' to the 2d gravity theory. 
We use this term in the following sense: In   equation  \eqref{eq:TotalAmp-QMIP} below the total amplitude for 2d gravity is expressed as an expectation value in an \emph{a priori} unrelated quantum 
mechanical system. This version of holography is more 
abstract than the usual picture of holography because the 
splitting space is not associated with any asymptotic boundary 
of spacetime. It should also be distinguished from the holographic 
dual as discussed in \cite{Marolf:2020xie}.

Closely related to $\bar\CA$ is the sum over all amplitudes
``from nothing to something.''
We will call this the \emph{Hartle-Hawking vector}
(abbreviated \emph{HH vector}), and it is defined by
\be
\Psi_{HH} : =  \oplus_{ X_o }  \bar\CA(\emptyset, X_o)(1)
\in  \CH_{\rm tot}  ~ . 
\ee
Our terminology is meant to distinguish this object 
from the ``Hartle-Hawking state'' 
which is a term sometimes used for a different thing in the quantum gravity literature. 
\footnote{It is worth remarking that sometimes a unitary TQFT $\CZ$ 
has $\CZ(\emptyset, X_o)(1) = 0$ for certain manifolds $X_o$. 
In this case the vector exists - it is the zero vector - but there is no corresponding quantum mechanical state for the quantum mechanical system with Hilbert space 
$\CZ(X_o)$. An example would be 3d $SU(2)$ Chern-Simons-Witten theory at a suitable level $k$ where $X_o$ is a suitable Lens space \cite{Freed:1991wd}. In this case the vector exists, although the state does not. We are not aware of any case where, 
after summing over bordisms the HH vector is zero. }
Similarly, we can define the \emph{Hartle-Hawking covector} 
as the amplitude for ``anything to go to nothing'': 
\be\label{eq:HHcovector-def}
\Psi_{HH}^\vee : =  \oplus_{ X_i  }  \bar\CA(X_i,\emptyset)  \in \Hom( \CH_{\rm tot} , \kappa):=\CH_{\rm tot}^\vee ~ . 
\ee
In addition to being conceptually important these objects will turn out to be 
computationally very useful in the 1d and 2d cases.

A remark that might be of general interest concerns ``global symmetries'' and 
continuous parameters. It is often said that there cannot be global symmetries in quantum gravity. It has also recently been argued that the absence of global symmetries  implies the absence of any free parameters in a quantum theory of gravity \cite{McNamara:2020uza}. In the 
ultra-simple,  very reduced, setting of the present paper these statements do not hold. First, as noted above, the seed TQFT's we will describe below typically come in continuous families, and the amplitude $\bar\CA_{\lambda}$ exists and is sensible for generic values of parameters $\lambda$. Regarding global symmetries, 
in the one-dimensional case discussed in section \ref{sec:OneD-TQFT} the seed TQFT will have $O(b)$ symmetry, where $b$ is the nondegenerate quadratic form used to define the theory. Then in the averaged theory the statespace $S^{\bullet} V$ is an $O(b)$ representation and the averaged amplitudes $\bar \CA$ are $O(b)$ equivariant. Similarly, in the two-dimensional case discussed  in section \ref{sec:TwoD-TQFT} an easy example of a theory with nontrivial global symmetry is provided by a semisimple Frobenius algebra with a symmetric group action permuting the idempotents and preserving the Frobenius trace. Then the averaged amplitudes are equivariant for this symmetric group action.
\footnote{In the context of ``non-factorizing theories'' of quantum gravity the existence of global symmetries and continuous parameters is well-known and has been explored in 
\cite{Iliesiu:2019lfc}, \cite{Kapec:2019ecr}, \cite{Chen:2020ojn}, \cite{Hsin:2020mfa}, \cite{Benini:2022hzx}. }
Of course, the very simple setting of the present paper does not exhibit important physical phenomena such as black holes, so we believe there is no sharp contradiction with the literature on quantum gravity.

\subsection{Relation To The Work Of Marolf-Maxfield}\label{subsec:RelateMM}
 
We now make some brief comments on the relation of 
our work to reference \cite{Marolf:2020xie}. 
\footnote{Our terminology and notation for 2d TQFT follows that of   \cite{Moore:2006dw}.}
Marolf and Maxfield begin their discussion 
by positing an action principle for a 2d 
``topological theory. '' The action is a function of the 
topological type of a compact surface $\Sigma$ 
with $g$ handles and a boundary that consists of $n$ disjoint circles
and is given by: 
\be\label{eq:MM-ident-1}
e^{-S} = e^{S_0\chi(\Sigma)+ n S_\partial} 
\ee
where $S_0, S_\partial$ are \emph{a priori} complex numbers 
parametrizing the theory 
and we recall that $\chi(\Sigma) = 2-2g -n$.  
It is problematic to interpret $n S_\partial$  
as a local term in an action and indeed 
there is some extended discussion in  \cite{Marolf:2020xie}
about suitable physical values for the parameter $S_{\partial}$. 
We will find a home for the parameter $S_\partial$ in a fully local 
2d TQFT, although it will not be a coupling constant in an action. 

In order to compare to our results the first question that 
must be addressed is the identification of the physical 
meaning of the $n$ boundary 
circles of $\Sigma$. In our framework there are two 
possible such identifications: 

\begin{enumerate} 

\item Ingoing or outgoing circles of a two-dimensional bordism. 
In that case the functor $\CZ$ assigns to circles factors of the 
ingoing or outgoing space of closed string states. 

\item Closed constrained boundaries, in the language of \cite{Moore:2006dw}.

\end{enumerate} 

The translation to our results depends on which identification
we adopt. We first attempt to identify ``amplitudes'' in the 
two discussions. Then we address the interpretations of these 
amplitudes.

In both identifications the Marolf-Maxfield model is related 
to the case where the seed TQFT has $\CC := \CZ(S^1)$ a one-dimensional 
Frobenius algebra. Denoting the idempotent generating $\CC$
by $\varepsilon$ the Frobenius trace is: 
\be\label{eq:MM-ident-2}
\theta(\varepsilon) = e^{\half S_0} ~ .
\ee 

Now, in the first identification, there is no discussion in 
 \cite{Marolf:2020xie} of the distinction between in-going and 
 out-going state spaces. The most natural way to reproduce their 
equation $(3.14)$ is to take all the circles to be ingoing and 
then identify this as the value of the ``Hartle-Hawking covector''
(see equation \eqref{eq:HHcovector-def} below) evaluated on 
the exponentiated diagonal 
\be\label{eq:MM-ident-3}
\exp[\tilde u \varepsilon] \in S^\bullet \CC 
\ee
where 
\be\label{eq:MM-ident-4} 
\tilde u = u_{\rm Marolf-Maxfield} e^{S_{\partial} - S_0}  ~ . 
\ee 
There is, of course, a similar interpretation if we take all 
the boundary circles to be out-going. 

If we adopt the second identification, and regard the circular 
boundaries as ``closed constrained boundaries,'' 
then we can identify the amplitude $(3.14)$ of  \cite{Marolf:2020xie}
as the sum over amplitudes for bordisms 
$\emptyset \rightarrow \emptyset$  
with $L$ closed constrained boundaries, weighted by fugacity $x^L$ and a combinatoric factor $1/L!$, and summed over $L$. 
In this identification all boundaries are endowed with a single boundary condition. In general, the boundary conditions in an open-closed 2d 
TQFT are labeled by objects in a category \cite{Moore:2006dw}. 
As shown in \cite{Moore:2006dw} in the semisimple case the 2d TQFT 
has an interpretation as a string theory with a  $0$-dimensional 
target space. In this case, the relevant category is the category of vector bundles over spacetime. If $\dim \CC = 1$ the target spacetime consists of a single point, so a vector bundle amounts to a choice of a single vector space $W$ of dimension $\dim W = w$. Our amplitude reproduces 
$(3.14)$ of  \cite{Marolf:2020xie} with the identification 
\be\label{eq:MM-ident-5} 
\frac{x}{\mu}w = u_{\rm Marolf-Maxfield} e^{S_{\partial} - S_0}  . 
\ee 
(Here $\mu^2 = \theta(\varepsilon)$, so $\mu^{-1}$ is the open string coupling constant.) 
In either identification there is no reason, from our viewpoint, 
to restrict the value of $S_{\partial}$, so long as it is compatible 
with equations \eqref{eq:MM-ident-4} or \eqref{eq:MM-ident-5}, respectively. 
 
Thus, in the second identification,
results of   \cite{Marolf:2020xie} are most naturally 
compared with equation \eqref{eq:ConstrainedVacVac} below. 
Equation \eqref{eq:ConstrainedVacVac} generalizes to 
the case where $\dim ~ \CC>1$ and allows different 
boundary conditions on different constrained boundaries.
Such generalizations were also considered in \cite{Gardiner:2020vjp}. 

Moving on to the case of open-closed TQFT, clearly we should attempt to connect to the discussion of ``EOW branes'' in \cite{Marolf:2020xie}.
It is natural to identify the locus of the EOW branes in a two-dimensional 
surface with the constrained boundaries that connect endpoints of 
in-going or out-going open string intervals. Fortuitously, these boundaries 
are denoted by dashed lines in both \cite{Marolf:2020xie} and \cite{Moore:2006dw}. 
The most natural identification proceeds by identifying the quantity 
$(\psi_j, \psi_i)$ in \cite{Marolf:2020xie} with a matrix unit 
$e_{ij} \in \CO_{aa}$ where $\CO_{aa}= \CZ(I_{aa})$ is the 
value of the TQFT on an open interval representing a morphism $a\to a$. 
\footnote{We must stress that in this paper constrained boundaries are labeled by 
vector \underline{spaces} and not by vectors in a vector space.}
To compare with \cite{Marolf:2020xie} take $\dim \CO_{aa} = k^2$ 
and consider  equation $(3.34)$ of \cite{Marolf:2020xie}.  Again to make a comparison to our results we must make a choice as to whether the open intervals are ingoing or outgoing. Choosing them all to be ingoing we are once again evaluating the Hartle-Hawking covector on the exponentiated diagonal in 
$S^\bullet \CO_{aa}$. Compare with our equations \eqref{eq:GardMeg-formula}
and \eqref{eq:SumConstrain2} below. The equations are close but 
not exactly the same because in our formula the determinant is raised to the power $1/\mu$. The origin of the discrepancy is the different handling of the ``open string amplitudes.'' (We weight them using the power of the open string coupling $\mu^{-1}$, as dictated by the rules of open-closed string theory.)

Having made a detailed identification of amplitudes we proceed to a 
comparison of their interpretations. For simplicity we restrict the discussion to the 
closed string case. Identifying the closed circle boundaries as 
closed constrained boundaries (identification two above) 
one can attempt to identify the amplitudes with the partition function of a topological quantum mechanics 
on $n$ disjoint circles. For this to work the target space category of the 1d TQFT should be the category of finite-dimensional vector bundles over measure spaces. We do not pursue this potential holographic interpretation in this paper. 

If, instead, we identify the closed circles as in-going or out-going 
boundaries in a two-dimensional bordism then the question arises as 
to the relation between $\CW$ and what is called the ``baby universe Hilbert space''
$\CH_{BU}$ in \cite{Marolf:2020xie}. In order to compare these it is useful 
to summarize the definition of $\CH_{BU}$ given in \cite{Marolf:2020xie} in 
the language of the present paper. (See also  \cite{Gesteau:2020wrk} for related remarks.) 
Since $\CH_{\rm tot}$ is a symmetric product, it has a natural algebra structure, 
the symmetric algebra structure. (This makes no use of the Frobenius structure on $\CC$ in 
the 2d case.) Assuming $\kappa = \IC$, one also provides $\CH_{\rm tot}$ with a real structure 
$K$, that is, $K$ is an antilinear involution, and thus $\CH_{\rm tot}$ becomes a $*$-algebra. 
The next step imitates the GNS construction of a representation of a $C^*$-algebra based on 
a state. 
\footnote{We say ``imitates'' because if
one wished to invoke the GNS construction one would first have to 
define a positive norm on $\CH_{\rm tot}$ (and complete with respect to this norm) 
such that the norm satisfies  the $C^*$ algebra 
condition $\parallel \phi K(\phi)\parallel = \parallel \phi\parallel^2$. 
We have not attempted to do this.} 
In the present case the state used to define the representation is the HH covector. 
Thus, one defines the sesquilinear form on the $*$-algebra $\CH_{\rm tot}$ by 
\footnote{If we put a Hilbert space structure on $\CH_{\rm tot}$ then this can 
also be defined as a matrix element of the total amplitude 
$\langle \phi_1, \phi_2 \rangle = (\phi_1^\dagger, \bar\CA(\phi_2))$ where 
the map $\phi\to \phi^\dagger\in \CH_{\rm tot}^\vee$ is defined by the Hilbert 
space structure and the pairing on the right is the natural dual pairing.}
\be\label{eq:BU-sesqui}
\langle \phi_1, \phi_2 \rangle := \Psi_{HH}^\vee(K(\phi_1) \phi_2) 
\ee
If (as is true in the $d=1,d=2$ cases studied below) the annihilator of the 
form is a linear space $N$ then, by definition, 
\be 
\CH_{BU} := \CH_{\rm tot}/N  ~ . 
\ee
The sesquilinear form \eqref{eq:BU-sesqui} descends to a positive definite
form on $\CH_{BU}$ and it is thus a Hausdorff pre-Hilbert space, which may be 
completed in the standard way. In this way one can compute, in the examples 
studied below, that $\CH_{BU}$ is an infinite-dimensional separable Hilbert space. 
Of course, all separable, infinite dimensional Hilbert spaces are isomorphic as 
Hilbert spaces, but there does not appear to be a natural isomorphism between 
$\CH_{BU}$ and $\CW$ in general. In other words, it is not obvious, at least not to us, 
why the space $\CH_{BU}$ automatically would serve as a splitting space. 
\footnote{It is also worth summarizing, in our language, the construction of 
section 3.3 of \cite{Marolf:2020xie}. Here $\CH_{\rm tot}$ is a symmetric 
algebra generated by $\varepsilon$ and $K(\varepsilon) = \varepsilon$. The 
state is defined by 
\be 
\Psi_{HH}^\vee(\varepsilon^n) = e^{\lambda}B_n(\lambda) 
\ee
and for $\lambda>0$ this is a nonnegative linear functional. It follows 
that if $f = \sum_{n=0}^\infty c_n \varepsilon^n \in \CH_{\rm tot}$ then 
\be 
\langle f_1, f_2 \rangle = \sum_{d=0}^\infty \frac{\lambda^d}{d!} \overline{f_1(d)} f_2(d) 
\ee 
where $f(x):=\sum_{n=0}^\infty c_n x^n$. Appendix A of  \cite{Marolf:2020xie} argues 
that the symmetric algebra (which formally only includes polynomials in $\varepsilon$) 
should be expanded to a certain space $V$ of entire functions $f(x)$ of order $\leq 1$. 
In this case the null space $N$ is the space of such entire functions that vanish on the 
nonnegative integers. Then there is a natural isomorphism 
\be 
V/N \cong \{ (\xi_0, \xi_1, \dots ) \in \IC^\infty \vert 
\sum_{d=0}^\infty \frac{\lambda^d}{d!}\vert \xi_d \vert^2 < \infty \} 
\ee 
given by $f\mapsto \{ \xi_d \}_{d=0}^\infty$ where $\xi_d := f(d)$. 
The sequence space is,  in turn, naturally isomorphic to the standard highest 
weight representation of the Heisenberg algebra. Thus, in the closed $d=2$ case 
there is a natural isomorphism of $\CW \cong \CH_{\rm tot} \cong \CH_{BU}$. } 
We close this discussion with two final comments on the relation to 
\cite{Marolf:2020xie}. First, an important motivating idea in \cite{Marolf:2020xie}
is that what we call the HH covector provides a set of expectation values,  for 
some hypothetical measure space,   for a stochastic variable on $\CH_{\rm tot}$,  denoted $Z(J)$ 
in \cite{Marolf:2020xie} where $J\in \CH_{\rm tot}$. Importantly, the measure space is a 
\underline{family} of quantum mechanical 
systems and $Z(J)$ also has the interpretation of a 
partition function of those ``holographically dual'' systems. We
will not be emphasizing this interpretation in the present paper.  
Rather we offer an alternative interpretation of the mathematical formulae in 
section \ref{subsubsec:RelationCoherent} below. 
Second, an  important distinction in terminology between our 
work and \cite{Marolf:2020xie} is that our ``Hartle-Hawking vector'' 
should not be confused with the ``Hartle-Hawking state,'' as 
used in \cite{Marolf:2020xie}. 

We now provide a dictionary of some key terms.  Since our notation varies for the different models we consider, we will employ the terminology used in the 2d closed semisimple case wherever possible. 

Note that the terminology in this paper is more in line with that of the TQFT literature. 

\clearpage 

\begin{center}
\begin{table}
\renewcommand{\arraystretch}{1.8}
\begin{tabular}{|p{8cm}|p{8cm}|}
\hline
This Paper & Previous Literature \\
\hline
\underline{Constrained boundaries}: These are worldlines of open string endpoints carrying the data of objects in a linear category $\fB$. It is also possible to include closed constrained boundaries labeled by objects in $\fB$. In the semi-simple case the objects correspond to finite rank vector bundles over the spectrum of the Frobenius algebra \cite{Moore:2006dw}. When the bordism category includes the data of closed constrained boundaries these should be summed over, as in \eqref{eq:SumConstrain1} and \eqref{eq:SumConstrain2}. 
& 
\underline{End-of-World Branes}: These are the worldlines of open string endpoints that carry a vector in a vector space. The object $(\psi_j, \psi_i)$ in \cite{Marolf:2020xie} is most naturally compared to $e_{ij} \in \CO_{aa}$. Equation $(3.34)$ of  \cite{Marolf:2020xie} and Equation $(4.27)$ of \cite{Gardiner:2020vjp} should be compared to 
\eqref{eq:GardMeg-formula} and \eqref{eq:SumConstrain2} of the present paper. \\
\hline
\underline{Ingoing and outgoing boundaries}: These carry the data of states in the Hilbert space of the 2d TQFT. Considering the direct sum over all in- and out-going closed one-dimensional manifolds we obtain the Fock spaces   $S^{\bullet}\mathcal{C}$ and $S^{\bullet}\mathcal{C}^{\vee}$ respectively. 
& 
\underline{Asymptotic boundaries}: These are be viewed as in-going or out-going boundaries 
in the far past or the far future in \cite{Marolf:2020xie} although for comparison with this 
paper they can also be considered to be closed constrained boundaries.  \\
\hline 
\underline{Total amplitude}: The total amplitude $\bar{\mathcal{A}}\in\text{Hom}(S^{\bullet} \mathcal{C},S^{\bullet}\mathcal{C})$ captures the amplitude from an arbitrary configuration of ingoing boundary components to an arbitrary configuration of outgoing boundary components. The amplitude is found to obey a splitting formula of the form, $\bar{\mathcal{A}}=\Phi_{\lambda}\Phi^{\dagger}_{\lambda}$. 
& 
\underline{The full quantum gravity path integral}: The total amplitude corresponds to the full quantum gravity path integral in a topological model of 2d gravity with arbitrary ingoing and outgoing states.     \\
\hline
\underline{Intermediate or Splitting space}: An infinite-dimensional inner product space $\mathcal{W}$ used to describe a splitting formula for the total amplitude obtained from summing over 2d bordisms with arbitrary but fixed boundary conditions. The splitting space and the splitting map are not unique, but the inner product is unique. When the ``seed'' TQFT satisfies a suitable positivity condition the inner product space admits a Hilbert space completion. 
& 
\underline{Baby Universe Hilbert Space}: Although this is not a direct analog of the splitting space, it shares some common features. The terminology goes back to the work of Coleman et. al. (\cite{Coleman:1988cy, Giddings:1988wv, Banks:1989zw, Banks:1988je}) on quantum gravity with Euclidean wormhole geometries. Baby universes are spatially closed, and therefore associated states are singlets under gauge symmetries. The amplitudes obtained from the sum over 2d geometries are expressed in terms of expectation values of Hermitian operators acting on a special class of states in a certain infinite-dimensional space known as the Baby Universe Hilbert space. \\
\hline 
\end{tabular}
\end{table}
\end{center}

\begin{center}
\begin{table}
\renewcommand{\arraystretch}{1.8}
\begin{tabular}{|p{8cm}|p{8cm}|}
\hline
This Paper & Previous Literature \\
\hline
\underline{Fock space on the TQFT}: In our view the (Hilbert space completion of) 
the Fock space  $S^{\bullet}\mathcal{C}$ would naturally be called the ``baby universe Hilbert space,'' but that terminology does not appear to be consistent with previous literature. (See the table above.) So we do not use it.  
& 
The Fock space $S^{\bullet}\mathcal{C}$ has been considered in 
\cite{Couch:2021wsm,deMelloKoch:2021lqp} in contexts related to 
those of the present paper. 
\\
\hline 
\underline{Coherent vector}: A vector $\Psi_{\theta}\in S^{\bullet}\mathcal{C}$ that we use to define the map $\Phi_{\theta}:S^{\bullet}\mathcal{C}\to \mathcal{W}$. If we view $S^\bullet \CC$ as a Fock space representation of a Heisenberg algebra then $\Psi_{\theta}$ 
is a coherent state in the standard sense.  
& 
\underline{Hartle-Hawking state}: A class of states in the ``baby universe Hilbert space'' 
indexed by a real parameter $\lambda$ analogous to the gravitational coupling. \\
\hline 
\underline{Hartle-Hawking vector}: A vector $\Psi_{\text{HH},\lambda}\in S^{\bullet}\mathcal{C}$ that describes a suitably weighted sum over all bordisms from the empty set to an arbitrary number of disconnected outgoing boundary components. Its dual version, the sum over ``everything to nothing,'' is the Hartle-Hawking covector. 
& \underline{No analogue:} The Hartle-Hawking vector and covector are not explicitly studied in the previous literature. However, subject to the remarks of section \ref{subsec:RelateMM} above, 
 they are implicitly used since the amplitudes presented in
\cite{Marolf:2020xie, Gardiner:2020vjp, Balasubramanian:2020jhl} are what we would call the Hartle-Hawking covector evaluated on the exponentiated diagonal. \\
\hline
\end{tabular}
\caption{Dictionary between some of the concepts in this paper and in \cite{Marolf:2020xie, Gardiner:2020vjp, Balasubramanian:2020jhl}}\label{Dictionary}
\end{table}
\end{center}

\section*{Acknowledgements}

We thank T. Banks, Y. Fan, D. Freed, S. Gukov, M. Hopkins, D. Jordan, M. Kontsevich, F. Luo, H. Maxfield, G. Segal, S.-H. Shao, and E. Witten for useful discussions and remarks. We are very grateful to T. Banks and H. Maxfield for comments on the 
draft. This work is supported by the US DOE   under grant
DOE-SC0010008 to Rutgers.

\section{One-Dimensional TQFT}\label{sec:OneD-TQFT}

We first consider the example of a $1$-dimensional TQFT. The data
defining the field theory is a pair $(V,b)$, where $V$ is a finite-dimensional vector space over
 $\kappa$ and $b$ is a nondegenerate symmetric form on $V$.

In the 1d case $X_i$ and $X_o$ will simply be disjoint unions of $n_i$ and $n_o$ points where $n_i, n_o$ are nonnegative integers. We can therefore denote the summed amplitude
in \eqref{eq:Sum-TQFT} by $\CA(n_i, n_o)$. As discussed near equation 
\eqref{eq:SymmetricDescent} the summed amplitude is a map
$\CA(n_i, n_o): V^{\otimes n_i} \to V^{\otimes n_o}$ but, thanks to the sum over 
bordisms, 
the map is totally symmetric and therefore descends to a map of symmetric products:
\be
\bar\CA(n_i, n_o) \in \Hom(S^{n_i}V , S^{n_o} V)  ~ . 
\ee
The ``total amplitude'' in this case is just an endomorphism of the full symmetric algebra:
\be
\bar\CA := \oplus_{n_i\geq 0, n_o \geq 0 }  \bar\CA(n_i, n_o) \in \Hom(S^{\bullet} V, S^{\bullet} V) ~ .
\ee

In the present case the HH vector and covector are 
defined by 
\be
\Psi_{HH} : =  \oplus_{ n_o \geq 0 }  \bar\CA(0, n_o)(1)  \in  S^{\bullet} V ~ . 
\ee
and
\be\label{eq:HHcovector-def}
\Psi_{HH}^\vee : =  \oplus_{ n_i \geq 0  }  \bar\CA(n_i,0)  \in \Hom( S^{\bullet} V, \kappa)
\ee
respectively. 

Note that if $\kappa = \IC$ and $b>0$ then we can endow $V$ with the structure of a Hilbert space. In that case we can identify
$S^\bullet V$ with a   Fock space generated by $V$.

The HH vector and covector are of conceptual importance, but are also
computationally very useful objects for the following reason.
The nondegenerate pairing $b: V\otimes V \to \kappa$ corresponds to the
``$U$-shaped bordism,'' the unique connected bordism with  2 ingoing points and empty outgoing boundary.
 The dual bordism  $\kappa \to V\otimes V $ is the value of the field theory
 on the $U$-shaped bordism from $\emptyset$ to two outgoing points.
  The nondegenerate pairing $b$ defines two maps
\be
\begin{split}
b^\vee: V \to V^\vee \\
b_\vee: V^\vee \to V ~ .  \\
\end{split}
\ee
To be explicit: $b^\vee(x)$ is that linear functional $\ell$ on $V$ so that $\ell(y) = b(x,y)$.
And $b_\vee(\ell)$ is that element $x \in V$ so that $\ell(y) = b(x,y)$. By applying $b_\vee$
to an ingoing boundary we can produce the amplitude for an outgoing boundary, and similarly,
$b^\vee$ takes an outgoing boundary to an ingoing boundary.  Therefore, once one has computed
either the HH vector or covector one can apply suitable powers of $b_\vee$ or $b^\vee $ to
obtain the general amplitude. See Appendix \ref{App:Apply-b} for a discussion of a small
subtlety when doing this.

We now evaluate the total amplitude.

We first consider the case $n_i = n_o=0$. That is, we consider
the  amplitudes for bordisms from $\emptyset$ to $\emptyset$.
These are just arbitrary disjoint unions of circles. The amplitude
for these is
\be
\CZ( \amalg S^1) = (\dim V)^c
\ee
where $c$ is the number of connected components of the bordism.
The amplitude for the empty bordism from $\emptyset$ to $\emptyset$ is just equal to $1$.
(This is true in every topological field theory.) So \eqref{eq:Sum-TQFT} becomes in
this case
\be
\bar\CA(0,0) = \exp[\dim V] := \Omega_V  ~ .
\ee
The exponentiation comes from the rule that we weight amplitudes by $\frac{1}{\vert \Aut(Y)\vert}$.
This is the standard relation between connected and disconnected diagrams in perturbation theory.

Before moving on to the case with nonempty in/out boundaries we  
begin with some linear algebra preliminaries. If $\kappa$ is not of 
characteristic two, as we will henceforth assume, there always
exists a decomposition of $V$ into  $b$-orthogonal lines:
\be
V = \oplus_{\alpha=1}^\nu V_\alpha   ~ .
\ee
In other words, there is a basis  $\{v_\alpha\}$ for $V$ so that
\be
b(v_\alpha, v_\beta) = b_\alpha  \delta_{\alpha, \beta}   ~ .
\ee
If $\kappa= \IR$ then we can choose $b_\alpha \in \{ \pm 1 \}$.  
If $\kappa = \IC$ then we can choose $b_\alpha = 1$, but we will 
leave $b_\alpha$ arbitrary for the moment to make clear basis-dependence 
of some of our formulae. 
In any case, for fixed $b_{\alpha}$, the basis 
$\{v_\alpha\}$ is not unique, but any two are related by the action of the 
orthogonal group $O(b)$. 
Finally, given   a choice of basis $\{ v_\alpha \}$ there is a unique dual basis of co-vectors $\check{v}^\alpha(v_\beta) = \delta^{\alpha}_{~\beta}$.  Note that
\be
b^\vee(v_\alpha) = b_\alpha \check{v}^\alpha \qquad \qquad b_\vee(\check{v}^\alpha) = \frac{1}{b_\alpha} v_\alpha  ~ ~ . 
\ee

Let us first consider the case where $V$ is one-dimensional, or, 
equivalently, let us 
restrict our considerations to the case where in and out states 
come from a single summand $V_\alpha$. 
It is then a straightforward matter to compute the HH vector and covector and
we obtain:
\be
\begin{split}
\Psi_{HH,\alpha}^\vee & = \Omega_{V_\alpha} \sum_{n=0}^\infty \frac{(2n)!}{n! 2^n}  b_\alpha^n (\check{v}^\alpha)^{2n} \\
\Psi_{HH,\alpha} & =\Omega_{V_\alpha}  \sum_{n=0}^\infty \frac{(2n)!}{n! 2^n}  b_\alpha^{-n}  (v_\alpha)^{2n} \\
\end{split}
\ee
where $\frac{(2n)!}{n! 2^n}$ is the number of ``Wick contractions'' decomposing the in or outgoing points into disjoint sets of pairs of points and $\Omega_{V_{\alpha}}= \exp[1]=2.718281828459....$.

\bigskip

\textbf{Remarks}:

\begin{enumerate}

\item As we have mentioned, it is important we only divide by automorphisms which are the identity on the boundary.
If we change the rule and divide by all automorphism then the $n^{th}$ term would be divided by $(2n)!$
and we can sum the series to get a different vector
 $\tilde \Psi_{HH, \alpha} = \Omega \exp[v_{\alpha}^2/(2 b_\alpha)] \in S^\bullet V$.
In either case the contractions
\be
( \Psi_{HH,\alpha}^\vee, \Psi_{HH,\alpha} )  \qquad\qquad
( \tilde\Psi_{HH,\alpha}^\vee, \tilde \Psi_{HH,\alpha} )
\ee
are \underline{not} the same as the vacuum to vacuum amplitude - a fact which some physicists
might find surprising. If we work with 
$\kappa=\IC$ and $b_\alpha>0$ and use the Hilbert space 
structure described in equation \eqref{eq:SesquiLinear-1d} below then $\Psi_{HH,\alpha}$ is not a normalizable vector.

\end{enumerate}

\bigskip

Having determined the HH covector the general amplitude then follows from
applying the operator $(b_\vee)^{n_{\rm in}}$ to get:
\footnote{Note that it is not really necessary to introduce a square root of $b_\alpha$, which might not exist for
a general field $\kappa$. The reason is that the net factor is   $(b_\alpha)^{n_i/2} (b_\alpha)^{-n_o/2} = b_\alpha^{n} b_\alpha^{-n_o}$.}
\be
\bar\CA_{\alpha} = \Omega_{V_\alpha}  \sum_{n_i + n_o=2n}   \frac{(2n)!}{n! 2^n}   (b_\alpha^{1/2} \check{v}^\alpha)^{n_i} \otimes (b_\alpha^{-1/2} v_\alpha)^{n_o} ~ . 
\ee

We can now say more precisely what we mean by the splitting formula \eqref{eq:SplittingFormula}.
At this point we need to choose $\kappa= \IR$ or $\kappa=\IC$, and for simplicity we will choose 
the latter. 
We first recall a basic fact related to Wick's theorem: The number of pairings
of $2n$ points is given by a Gaussian integral:
\be
\frac{(2n)!}{n! 2^n } = \int_{-\infty}^{+\infty}  \frac{dh}{\sqrt{2\pi}} e^{-\half h^2} h^{2n}
\ee
So we define a linear map of complex vector spaces: 
\be
\Phi_{b_{\alpha}}: S^\bullet V_\alpha \rightarrow  L^2(\IR)
\ee
by the formula:
\be
\Phi_{b_{\alpha}} := \sqrt{\Omega_{V_\alpha}}
\sum_{n}   (b_\alpha^{1/2} \check{v}^\alpha)^{n} \otimes \psi_n
\ee
where $\psi_n \in L^2(\IR)$ is the vector with 
\be 
\psi_n(h) = (2\pi)^{-1/4} h^n e^{-\frac{1}{4}h^2} ~ . 
\ee

To define $\Phi_{b_{\alpha}}^\dagger$ we need a nondegenerate inner 
product.  The basis $\{ v_\alpha\}$ canonically determines a
real structure on $V$ and we can use this to define a sesquilinear form
\footnote{Note that for $\kappa=\IC$ this 
choice of sesquilinear form breaks the $O(b)$-symmetry from a complex orthogonal 
group to a real orthogonal group.}
\be\label{eq:SesquiLinear-1d}
h(z_\alpha v_\alpha, w_\beta v_\beta ) := 
\langle z_\alpha v_\alpha, w_\beta v_\beta\rangle:= \sum_\alpha b_\alpha  z_\alpha^* w_\alpha
\ee
Such a nondegenerate 
form defines an anti-linear isomorphism $V \to V^\vee$. 
(In Dirac's bra-cket notation this is nothing but the map 
 $\vert \psi \rangle \rightarrow \langle \psi \vert$. We will 
 denote it by $v\to v^\dagger$.) Then we have 
\be 
(v_\alpha)^\dagger = b_\alpha \check{v}^\alpha  \qquad\qquad  ( \check{v}^\alpha )^\dagger = (b_\alpha^*)^{-1} v_\alpha ~~ .
\ee
%
%Because of the complex conjugation we will need to use %$\Phi_{b_\alpha^*}^\dagger$ where the $\dagger$ is 
%taken with respect to the complex conjugated bilinear form. 
%

Similarly, the   sesquilinear form on $L^2(\IR)$: 
\be 
\langle\psi_1, \psi_2 \rangle = \int_{-\infty}^{+\infty} dh \psi_1(h)^* \psi_2(h)
\ee
defines a canonical anti-linear isomorphism $L^2(\IR) \to L^2(\IR)^\vee$. 
To write a splitting formula we need to choose a basis with $b_{\alpha}$ real and then: 

\be\label{eq:AnnoyingPhase}
\Phi_{b_\alpha}  \Phi_{b_{\alpha}}^\dagger 
 = \Omega_{V_\alpha} \sum_{n_i + n_o=2n}   \frac{(2n)!}{n! 2^n}   
 (b_\alpha^{1/2} \check{v}^\alpha)^{n_i} \otimes   (b_\alpha^{-1/2} v_\alpha)^{n_o}
\ee
In other words: 
\be\label{eq:1d-splittingformula}
\bar\CA_{\alpha} = \Phi_{b_\alpha} \Phi_{b_{\alpha}}^\dagger  ~ .
\ee

We now generalize the above remarks to the case $\dim V>1$. Since the 
amplitude must be $O(b)$ invariant we can immediately write down the 
HH vector and co-vector given the one-dimensional result: 
\be\label{eq:HH-GenV} 
\Psi_{HH} = \Omega_V \sum_{n=0}^\infty  \frac{(2n)!}{n!2^n} (v^2)^n 
\in S^{\bullet} V 
\ee 
where $v^2\in S^2 V$ (which should not be confused with the scalar quantity $b(v,v)\in \kappa$) is defined by the $O(b)$ invariant expression: 
\footnote{To check $O(b)$ invariance note that a change of basis $v_\alpha \to A_{\beta \alpha} v_\beta$ is in $O(b)$ if
$A^{tr} D(b) A = D(b)$ where $D(b)_{\alpha\beta} = b_\alpha \delta_{\alpha, \beta}$. Now use 
the formula $D(b)^{-1} = A D(b)^{-1} A^{tr}$. In a general basis we have 
$v^2 = v_\alpha Q_{\alpha, \beta}   v_\beta$ where $Q_{\alpha, \beta}$ is the 
matrix inverse of $b(v_{\alpha} , v_{\beta})$. } 
\be
v^2 := \sum_\alpha b_\alpha^{-1} v_\alpha v_\alpha \in S^2 V  ~  .
\ee

Note that this can be written as 
\be\label{eq:HH-GenV-2} 
\Psi_{HH} = \Omega_V \sum_{\vec\ell\in \IZ_+^\nu} c(2\vec \ell)
 \prod_{\alpha=1}^\nu
\frac{(2\ell_\alpha)!}{\ell_\alpha! 2^{\ell_\alpha}} (b_\alpha^{-1} v_\alpha^2)^{\ell_\alpha}
\ee 
where $c(2\vec \ell)$ is just the multinomial 
\be 
c(2\vec \ell):= {2n\choose 2\ell_1, 2\ell_2, \dots, 2 \ell_\nu} 
:= \frac{(2n)!}{(2\ell_1)! \cdots (2\ell_\nu)!}
\ee 
and  $n=\ell_1 + \cdots + \ell_\nu$. Equation \eqref{eq:HH-GenV-2} teaches us that  although $S^\bullet V \cong \otimes_\alpha S^\bullet V_\alpha$ it is 
\underline{not} true that $\bar\CA$ is the same as $\otimes_\alpha \bar \CA_{\alpha}$ because $c(2\vec \ell)$ monomials in the tensor 
algebra all project to the same monomial in the symmetric algebra. 

Introducing Gaussian integrals for the numbers of pairings and applying powers of $b^\vee$ we obtain the formula for the general amplitude: 
\be 
\bar\CA = \sum_{\vec\ell, \vec s \in \IZ_+^\nu} c(\vec \ell) c(\vec s)
\int \prod_{\alpha=1}^\nu \frac{dh_\alpha}{\sqrt{2\pi}}e^{-\half h_\alpha^2} (b_\alpha^{1/2} h_\alpha \check{v}^\alpha)^{\ell_\alpha} (b_\alpha^{-1/2} h_\alpha v_\alpha)^{s_\alpha} 
\ee
where $\nu = \dim V$ and   again 
\be 
\begin{split} 
c(\vec \ell) & :=  {n_i \choose \ell_1, \dots, \ell_\nu}
:= \frac{n_i!}{\ell_1! \cdots \ell_\nu!} \\ 
c(\vec s) & := {n_o \choose s_1, \dots, s_\nu} 
:= \frac{n_o!}{s_1! \cdots s_\nu!} \\ 
\end{split}
\ee
where $n_i = \ell_1 + \cdots + \ell_\nu$ and $n_o= s_1 + \cdots + s_\nu$. 

To write the splitting formula we introduce 
\be 
\Phi_b: S^\bullet V \to L^2(\IR^\nu) 
\ee 
defined by 
\be 
\Phi_b := \sqrt{\Omega_V} \sum_{\vec \ell \in \IZ_+^\nu} c(\vec \ell) 
\prod_{\alpha} (b_\alpha^{1/2} \check{v}^\alpha)^{\ell_\alpha} \otimes 
\psi_{\vec \ell }
\ee
where 
\be 
\psi_{\vec \ell} := (2\pi)^{-\nu/4} \prod_{\alpha} h_\alpha^\ell e^{-\frac{1}{4} h_\alpha^2} 
\ee
Once again, the splitting formula requires that all the $b_\alpha$ should be real and then 
\be 
\bar\CA = \Phi_{b} \Phi_b^\dagger ~ . 
\ee

\bigskip
\begin{enumerate}

\item    A natural question (posed to us by Mike Hopkins) is the following:
How are $\bar\CA(X,Y)$, $\bar\CA(Y,Z)$ and $\bar\CA(X,Z)$ related? In particular is
$\bar\CA(Y,Z)\circ\bar\CA(X,Y)$ the same as $\bar\CA(X,Z)$ as one might expect if there
were a unique morphism in each hom space of a category.
It is now evident that this is not at all the case for our amplitudes. In the case that $V$ is one-dimensional we can say 
\be\label{eq:Composition-1d}
\bar\CA(n_2, n_3) \circ \bar\CA(n_1,n_2) = f_{n_1,n_2,n_3} \bar\CA(n_1,n_3)
\ee
with a rather unpleasant fudge factor. Note that the ``cocycle'' $f_{n_1, n_2, n_3}$ is trivializable so that if we define
\be
\tilde \CA(n_1, n_2) = \frac{2^n n!}{(2n)!} \bar\CA(n_1, n_2)
\ee
then the $\tilde\CA(n_1, n_2)$ indeed compose coherently:
\be
\tilde \CA(n_2, n_3) \circ \tilde \CA(n_1,n_2) =  \tilde \CA(n_1,n_3)
\ee
%
%
%\be
%f_{n_1,n_2,n_3} = 2^{-n_2}  \frac{ (n_1+n_2)!(n_2 + n_3)!}{(n_1+n_3)!}
%  \frac{\left( \frac{n_1+n_3}{2}\right)! }{ \left( \frac{n_1+n_2}{2}\right)! \left(\frac{n_2 + n_3}{2}\right)!}
%\ee
%
but this does not appear to generalize easily to $\dim V > 1$.

\item Applying the remark from Appendix \ref{App:MultiLinear-Diagonal}
we learn that the general amplitude is
implicitly encoded in the value of the HH covector on the 
exponentiated diagonal: 
\be\label{eq:1d-exp-diag}
\Psi_{HH}^\vee(e^v) = \Omega_V \exp\left[  \frac{b(v,v)}{2} \right] ~ . 
\ee
 In our view, it is a mistake to try to write a splitting formula directly for the HH covector evaluated on the exponentiated diagonal.

\end{enumerate}

\section{Two-Dimensional Closed TQFT}\label{sec:TwoD-TQFT}

We now turn our attention to the case $d=2$. It is useful to begin with
the case where the objects in the bordism category are \underline{closed}
manifolds.
\footnote{Again, we remind the reader that for 2d TQFT we  follow the notation and terminology of \cite{Moore:2006dw}.}
In this case, as is very well-known, the TQFT is
given by a finite-dimensional commutative associative Frobenius algebra
over $\kappa$. We have the basic defining data of a
 commutative associative Frobenius algebra:
\be
\begin{split}
m: \CC \otimes \CC & \to \CC \\
\theta: \CC \to \kappa \\
\end{split}
\ee
where $m$ is the multiplication and $\theta$ is the trace and
$b:=\theta \circ m$ is a symmetric nondegenerate bilinear form.

The objects of the bordism category are now disjoint unions of circles.
Once again the objects of the bordism category are labeled by  nonnegative integers. So the sum over bordisms can be
denoted:
\be
\bar\CA(n_{\rm in}, n_{\rm out} ) := \sum_{Y: n_{\rm in}\to n_{\rm out} }  \frac{1}{\vert\Aut(Y) \vert} \CZ(n_{\rm in}, n_{\rm out})
\in  \Hom(S^{n_{\rm in}}\CC , S^{n_{\rm out}}\CC  )
\ee
and similarly
\be
\bar\CA \in  \Hom(S^{\bullet}\CC , S^{\bullet}\CC  )
\ee
is the direct sum over all $n_{\rm in}, n_{\rm out}  \geq 0$. We now evaluate these amplitudes in various special
cases in terms of the defining data of the Frobenius algebra.

\subsection{The Vacuum To Vacuum Amplitude}

The special case of $n_{\rm in} = n_{\rm out} =0 $ can be evaluated easily for any 2d TQFT:
We introduce a basis  $\{ e_x \}$ for $\CC$  together with a
dual basis of vectors $e^x \in \CC$  so $\theta(e^x e_y ) = \delta_{x,y}$.
Then we can define the  handle-adding element
\be
\fh:= \sum_x  e^x e_x \in \CC
\ee
It is then straightforward to see that the
 sum over nonempty connected bordisms $\emptyset \to \emptyset$ is
$ \theta(\frac{1}{1-\fh})$. As we mentioned in the introduction, the amplitude
exists for the generic theory, but if the endomorphism   $L(\fh):\CC\to \CC$
defined by   multiplication with $\fh$ has eigenvalue $1$, then the sum diverges.

The sum over \underline{all} bordisms $Y$, weighted
by $\frac{1}{\vert Aut(Y)\vert}$ is
\be
\bar\CA(0,0) = \exp[ \theta(\frac{1}{1-\fh})] ~ .
\ee
Recall, again that  the empty bordism $\emptyset \to \emptyset$
has  amplitude $1$.

\subsection{The Semisimple Case}

In this case we can choose a basis of idempotents $e_x = \varepsilon_x$ with
\be
\varepsilon_x \varepsilon_y = \delta_{x,y} \varepsilon_x
\ee
and defining
\be
\theta_x:=  \theta(\varepsilon_x)
\ee
we have $e^x = \theta_x^{-1} \varepsilon_x$, so
\be
\fh = \sum_x  \theta_x^{-1} \varepsilon_x  ~ .
\ee

We   have an orthogonal decomposition
\be\label{eq:Decomp-x}
\CC \cong \oplus_x  \CC_x
\ee
so that $\CC_x \CC_y = 0 $ for $x\not= y$.
In the semisimple case each of the $\CC_x$ is the one dimensional image of
multiplication by the idempotent $\varepsilon_x$.
\footnote{In the interpretation of the 2d TQFT theory as a 0-dimensional
topological string theory the points $x \in {\rm Spec}(\CC)$ are disjiont ``universes''
in the target space of the string.}
 As we have noted, 
the amplitudes descend to the symmetric product, and
given \eqref{eq:Decomp-x} we have
\be\label{eq:TensorProd-Decomp}
S^\bullet \CC \cong \otimes_x  S^\bullet \CC_x ~ . 
\ee

Now, suppose we have a \underline{connected} bordism, and
on each of the input circles we have ``homogeneous'' input vectors,
i.e. input vectors with a definite component in the decomposition
\eqref{eq:Decomp-x}. Then, the amplitude is zero unless all the inputs
are in the \underline{same} component. That is, if we have $\varepsilon_x$
and $\varepsilon_y$ with $x\not=y$ as two inputs on a connected bordism
then the amplitude is zero. Moreover, for a connected bordism, all of
whose inputs are in a single summand $\CC_x$, the output will be in
$T^{\bullet}\CC_x$. This motivates us to consider first the 
amplitude $\CA_x$ in the case $\CC$ is one dimensional. 

Assuming now $\CC$ is one-dimensional it will be  generated
by a single idempotent $\varepsilon$. We will denote $\theta(\varepsilon)$ simply by $\theta \in \kappa$. 
We will begin by computing the Hartle-Hawking covector. Then
we will apply the $U$-bordism, which is represented by the
map $b_\vee$. Denoting the dual basis vector of $\CC^\vee$
by $\varepsilon^\vee$ we have:
\be\label{eq:OneDimDualization}
\begin{split}
b^\vee(\varepsilon) & = \theta \varepsilon^\vee \\
b_{\vee}(\varepsilon^\vee) & = \theta^{-1} \varepsilon\\
\end{split}
\ee

Now, for a connected morphism $\amalg S^1 \to \emptyset$ the
amplitude is easily evaluated to be
\be
\CZ(z_1 \varepsilon, \dots, z_n \varepsilon)= \prod_{i=1}^n z_i  \theta(\fh^p \varepsilon) = \theta^{1-p}  \prod_{i=1}^n z_i
\ee
where the surface has $p$ handles. Therefore summing over genera gives
\be
\bar\CA(n,0)^{\rm connected}(z_1 \varepsilon, \dots, z_n \varepsilon)= \lambda \prod_{i=1}^n z_i
\ee
where
\be
\lambda:= \frac{\theta}{1-\theta^{-1}}
\ee
\noindent\rule[0.5ex]{\linewidth}{1pt}
\usetikzlibrary{shapes.misc}

\tikzset{cross/.style={cross out, draw=black, minimum size=2*(#1-\pgflinewidth), inner sep=1pt, outer sep=1pt},
cross/.default={3pt}}
\begin{tikzpicture}[tqft/cobordism/.style={draw},tqft/every boundary component/.style={dashed}]
\pic [tqft/cap, name=a1, anchor={(0,2)}];
\draw (-4.5,-2) node[cross,black, rotate=45] {};
\draw (0.5,-2) node[cross,black, rotate=45] {};
\draw (3.5,-2) node[cross,black, rotate=45] {};
\draw (4.5,-2) node{\ldots} (5,-2);
\draw (5,-1.9)--(5.4,-1.9);
\draw (5,-2.1)--(5.4,-2.1);
\pic [tqft/cap, black, name=a4, fill=gray, anchor={(-2,0)}];
\pic [tqft/cup, black, name=b4, fill=gray, at=(a4-outgoing boundary 1)];
\pic [tqft/cap, name=a2, anchor={(2,1)}];
\pic [tqft/cap, name=a3, anchor={(4,0)}];
\pic [tqft/cup, name=b3, at=(a3-outgoing boundary 1)];
\pic [tqft/pair of pants, name=b2, at=(a2-outgoing boundary 1)];
\pic [tqft/reverse pair of pants, name=c2, at=(b2-outgoing boundary 1)];
\pic [tqft/cup, name=d2, at=(c2-outgoing boundary 1)];
\pic [tqft/pair of pants, name=b1, at=(a1-outgoing boundary 1)];
\pic [tqft/reverse pair of pants, name=c1, at=(b1-outgoing boundary 1)];
\pic [tqft/pair of pants, name=d1, at=(c1-outgoing boundary 1)];
\pic [tqft/reverse pair of pants, name=e1, at=(d1-outgoing boundary 1)];
\pic [tqft/cup, name=f1, at=(e1-outgoing boundary 1)];
\end{tikzpicture}

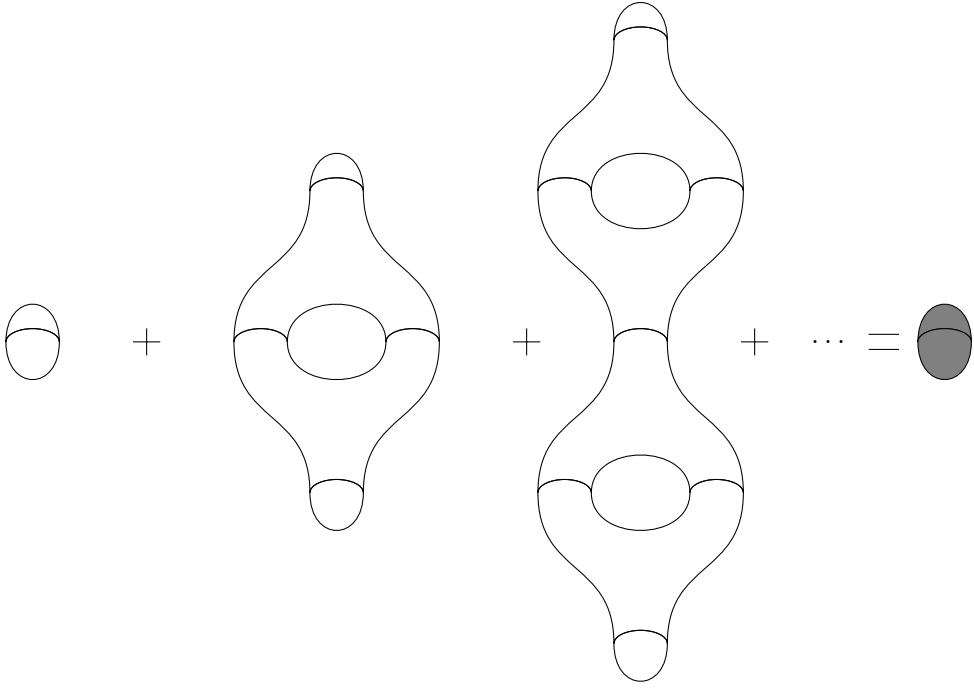
\textbf{Figure 2:} \emph{The definition of $\lambda$ as the sum of connected bordisms $X:\emptyset \to \emptyset$ with an arbitrary number of handles (we will henceforth use shading as in the bordism on the right hand side to indicate such a sum over all connected bordisms)}
\\
\noindent\rule[0.5ex]{\linewidth}{1pt}
Thus far, the considerations are pretty trivial. 
Something interesting happens when
we   include \underline{disconnected} surfaces 
(and this is the main insight of   Marolf and Maxfield  
\cite{Marolf:2020xie}). 
We consider partitions of the set of
incoming states into disjoint sets associated with the connected bordisms.
Thus we need to know how many ways there are of
dividing the sum over bordisms into $k_1$ connected components with one ingoing circle,
$k_2$ components with two ingoing circles and so forth,  so that we have $k_j$ components with  $j$ ingoing circles
so that $\sum_j j k_j$ is the total number of ingoing circles. For any number of ingoing circles the sum over
connected bordisms to $\emptyset$ gives a   \underline{single} factor of $\lambda$. Thus, with $n$ ingoing
circles the sum over all bordisms to the emptyset is  $B_n(\lambda)$,
where $B_n$ is the $n^{th}$ exponential Bell polynomial $B_n(x_1, x_2, \dots, x_n)$ evaluated for $x_1=x_2=\cdots= x_n=\lambda$.
  See Appendix \ref{App:BellPolynomials} (or Wikipedia)
for some basic facts about Bell polynomials. Including the disconnected
components with no boundaries gives an overall factor of $e^{\lambda}$. Altogether
we get
\be
\bar\CA(n,0)(z_1 \varepsilon, \dots, z_n \varepsilon)  = e^\lambda B_n(\lambda) \prod_i z_i  ~ . 
\ee

\textbf{Remarks}:

\begin{enumerate}

\item Note well that in our computation we did \underline{not} divide by the number of automorphisms of the
components of the surface. Thus if we have $k_1$ ingoing connected  with one ingoing circle
we do \underline{not} include an extra $1/k_1!$. The justification for this is that we are only considering
automorphisms of the surface which reduce to the identity on the boundary.

\item Next applying the $U$-morphism we immediately deduce
\be\label{eq:SimpGenAmp}
\bar\CA(n_i, n_o) = e^\lambda B_{n_i + n_o}(\lambda) \theta^{-n_o} (\varepsilon^\vee)^{n_i} \otimes \varepsilon^{n_o}
\ee
Note that, as a check on our factors:
\be
(b^\vee)^{n_o} (b_\vee)^{n_i} \bar\CA(n_i, n_o) = \bar\CA(n_o,n_i)
\ee
as expected.

\item
Returning to the question surrounding \eqref{eq:Composition-1d} 
when $\dim \CC =1 $ we compute
\be
\bar\CA(n_2, n_3) \circ \bar\CA(n_1, n_2) = f_{n_1,n_2,n_3} \bar\CA(n_1, n_3)
\ee
%
%
%   e^\lambda \frac{B_{n_1 + n_2}(\lambda) B_{n_2 + n_3}(\lambda)}{B_{n_1 + n_3}(\lambda)} \theta^{-n_2}
%
and the ``cocycle''  $f_{n_1,n_2,n_3}$ is trivializable so that if we
define 
\be 
\tilde\CA(n_1,n_2):= e^{\lambda} B_{n_1+n_2}(\lambda) \bar\CA(n_1,n_2)
\ee 
then
\be
\tilde \CA(n_2, n_3) \circ \tilde \CA(n_1, n_2) =   \tilde \CA(n_1, n_3)
\ee
When $\dim \CC >1$ this statement does not generalize naturally.

\item  We can now write down the Hartle-Hawking vector and
covector
\be
\begin{split}
\Psi_{HH} & = e^\lambda \sum_{n=0}^\infty B_n(\lambda)  \left( \frac{\varepsilon}{\theta} \right)^n \in S^\bullet \CC \\
\Psi_{HH}^\vee & = e^\lambda \sum_{n=0}^\infty B_n(\lambda)  \left(  \varepsilon^\vee \right)^n \in S^\bullet \CC^\vee \\
\end{split}
\ee
Note in particular that   the pairing of the dual $\Psi_{HH}^\vee$ with
$\Psi_{HH}$ is a formal series
\be\label{eq:SumBellSq}
e^{2\lambda} \sum_{n=0}^\infty \theta^{-n} B_n(\lambda)^2
\ee
for fixed $\lambda$, $B_n(\lambda)$ grows,
roughly speaking, like $n!$  \cite{BellAsymptotics} so the series is at best asymptotic.
\footnote{Note that 
since $B_n(\lambda)^2 = \lambda^{2n} + \cdots $ it is useful to consider the series 
\be\label{eq:SumBellSq-2}
\sum_{n=0}^\infty \theta^{-n} \frac{B_n(\lambda)^2}{(2n)!} 
\ee
Although \eqref{eq:SumBellSq-2}  is not exactly the Borel transform of 
\eqref{eq:SumBellSq}, we expect it will  have the same convergence properties. 
Using  theorem 2.1 of  \cite{BellAsymptotics} one can check that 
\eqref{eq:SumBellSq-2} converges to an entire function of $\lambda$. }
In particular, rather surprisingly,   $(\Psi_{HH}^\vee, \Psi_{HH})$ is
\underline{not} the vacuum to vacuum to vacuum amplitude! One should compare
this result with equation (2.7) of \cite{Marolf:2020xie}.

\item To compare to the previous literature, we again recall the
remark from Appendix \ref{App:MultiLinear-Diagonal}.    If we write
$\phi =\sum_x  u_x \varepsilon_x$ with $u_x \in \kappa$ then the value 
of the HH covector on the exponentiated diagonal is: 
\be
\Psi_{HH}^\vee(e^\phi) = \exp[ \sum_x  \lambda_x e^{u_x} ]
\ee
which, subject to the remarks of section \ref{subsec:RelateMM} above, 
can be compared with expressions in \cite{Gardiner:2020vjp}.

\end{enumerate}

\subsubsection{A Splitting Formula}

We now explain a curious splitting formula for the total amplitude.
This follows from the elementary expression for the Bell polynomials:
\be\label{eq:Bell-Expansion}
e^\lambda B_n(\lambda) =    \sum_{d=0}^\infty \frac{\lambda^d}{d!} d^n ~~ .
\ee
This formula, which is related to coherent states in the harmonic oscillator,
plays a central role in the discussion of \cite{Marolf:2020xie}.

 Combining equation \eqref{eq:Bell-Expansion} 
with equation \eqref{eq:SimpGenAmp} gives the full amplitude:
\be\label{eq:FullOneDimC}
\begin{split}
\bar\CA  & =\sum_{n_i, n_o \geq 0} (\varepsilon^\vee )^{\otimes n_i} \left( \sum_{d=0}^\infty \frac{\lambda^d}{d!} d^{n_i + n_o} \right)
\otimes \left( \frac{\varepsilon}{\theta} \right)^{\otimes n_o}    \\
\end{split}
\ee

To write the splitting formula it is useful to note that there is
a nondegenerate bilinear pairing on $S^\bullet \CC$ induced by the Frobenius structure on $\CC$.
We introduce the   ``orthonormal basis''  $\varepsilon/\sqrt{\theta}$ and $\sqrt{\theta} \varepsilon^\vee$.
This is the unique pair of vectors which are mapped into each other under $b^\vee$ and $b_\vee$  and also pair to one.
They induce an analogous ``orthonormal basis'' on the  symmetric product $ \{  \left( \frac{\varepsilon}{\sqrt{\theta}}\right)^{\otimes d} \} $
for $S^\bullet(\CC)$ and
and $ \{  \left( \sqrt{\theta} \varepsilon^\vee \right)^{\otimes d} \} $ for $S^\bullet(\CC^\vee)$.
Now we define 
\be 
\Phi_{\theta}: S^\bullet \CC  \to S^\bullet \CC 
\ee
by the formula 
\be 
\Phi_{\theta} = \sum_{n,d\geq 0} \sqrt{\frac{\lambda^d}{d!}} d^n 
 (\varepsilon^\vee)^{n} \otimes \left( \frac{\varepsilon}{\sqrt{\theta}} \right)^d
\ee
Then the natural dualization determined by the Frobenius structure gives 
\be
\tilde\Phi_{\theta} =  \sum_{n,d\geq 0} \sqrt{\frac{\lambda^d}{d!}} d^n
\left( \sqrt{\theta} \varepsilon^\vee \right)^d \otimes
 \left(\frac{\varepsilon}{\theta}\right)^{n} 
\ee
In  $\Phi_\theta \otimes \tilde \Phi_\theta$ we can contract the middle two factors of $S^\bullet \CC \otimes S^\bullet \CC^\vee \to \kappa$
using the natural dual pairing
to get an element of $S^\bullet \CC^\vee \otimes S^\bullet \CC \cong   \Hom(S^\bullet \CC, S^\bullet \CC)$ and
with that understood we have the result that, for a one-dimensional Frobenius algebra
\be\label{eq:OneDimC-Split}
\bar\CA =  \Phi_\theta \tilde\Phi_\theta ~ . 
\ee

\subsubsection{Generalization To $\CC$ with $\dim \  \CC>1$. }

We can now easily generalize equation \eqref{eq:FullOneDimC}
to the case $\dim \  \CC=n>1$ by noting that in the descent 
\eqref{eq:SymmetricDescent}, if $s_i$ are nonnegative integers then 
$c(\vec s)$  
distinct monomials in the tensor algebra descend to the 
same element $\prod_x \varepsilon_x^{s_x}$ in the symmetric 
algebra. Accordingly, the HH vector is 
\be 
\begin{split} 
\Psi_{HH} & = \sum_{\vec s \in \IZ_+^n}  c(\vec s) 
\prod_x \left( \sum_{d_x=0}^\infty \frac{\lambda_x^{d_x}}{d_x!} \left(d_x \frac{\varepsilon_x}{\theta_x}\right)^{s_x} \right) \\ 
& = \sum_{\vec d\in \IZ_+^n  } \prod_x \frac{\lambda_x^{d_x}}{d_x!} \sum_{m=0}^\infty\left( \sum_{x=1}^n d_x \frac{\varepsilon_x}{\theta_x}\right)^m \\
\end{split}
\ee
where $n = \dim \CC$. 
The full amplitude is 
\be\label{eq:2d-SS-Splitting-1} 
\begin{split} 
\bar\CA = \sum_{\vec \ell, \vec s \in \IZ_+^n} c(\vec \ell) c(\vec s)  
&
 \prod_x \left(  \sum_{d_x=0}^\infty (d_x \varepsilon_x^\vee)^{\ell_x} 
\frac{\lambda_x^{d_x}}{d_x!} \left( d_x \frac{\varepsilon_x}{\theta_x}\right)^{s_x}\right) \\ 
\end{split} 
\ee 
with splitting by $\Phi: S^\bullet \CC \to S^\bullet\CC$ given by 
the generalization of \eqref{eq:OneDimC-Split}: 
\be\label{eq:2d-SS-Splitting-2}
\bar\CA = \Phi_{\theta} \tilde\Phi_{\theta}
\ee
with
\be\label{eq:2d-SS-Splitting-3}
\Phi_{\theta} = \sum_{\vec \ell, \vec d \in \IZ_+^n} c(\vec \ell) 
\prod_{x} \sqrt{\frac{\lambda_x^{d_x}}{d_x!}} (d_x \varepsilon_x^\vee)^{\ell_x} \otimes \left( \frac{\varepsilon_x}{\sqrt{\theta_x}}\right)^{d_x}  ~ . 
\ee 

\subsubsection{Adjoint Splitting}

The expressions \eqref{eq:2d-SS-Splitting-1}- \eqref{eq:2d-SS-Splitting-3}
are valid over any field $\kappa$
(such that $\sqrt{\theta_x}$ and $\sqrt{\lambda_x^d/d!}$ make sense). 
When we take $\kappa=\IC$ we can introduce a sesquilinear form and 
write a slightly different splitting formula.

For $\kappa = \IC$ and $\CC$ semisimple there is the preferred
basis of the complex vector space $\CC$ given by the idempotents
$\varepsilon_x$. There is therefore a natural \underline{real structure}
(i.e. an anti-linear map squaring to one) defined by
\be
K: \phi = \sum_x u_x \varepsilon_x \mapsto \sum_x u_x^* \varepsilon_x ~ . 
\ee
Related to this, we can define a sesquilinear form on $\CC$ 
\be 
h(\sum_x u_x \varepsilon_x, \sum_y v_y \varepsilon_y) := \sum_x  u_x^* v_x \theta_x 
\ee
The sesquilinear form defines an anti-linear isomorphism $\dagger$ between $\CC$ and $\CC^\vee$ 
and on basis vectors it is: 
\be 
(\varepsilon_x)^\dagger = \theta_x \varepsilon_x^\vee  \qquad \qquad (\varepsilon_x^\vee)^\dagger = (\theta_x^*)^{-1} \varepsilon_x ~ . 
\ee

The Fock space  $S^\bullet \CC$ inherits a sesquilinear form 
\be
h(\varepsilon_I, \varepsilon_J)
=\langle \varepsilon_I, \varepsilon_J \rangle 
=  \prod_{i=1}^n \theta_{x_i} \delta_{I,J}
\ee
where $I=(x_1, \dots, x_n)$ is an unordered tuple and 
\be
\varepsilon_I = \prod_i  \varepsilon_{x_i} \in S^n \CC
\ee
We can now try to split the total amplitude by  
\be\label{eq:AlmostSplit-1}
\bar\CA {\buildrel ? \over = }   \Phi_\theta \Phi_{\theta^*}^\dagger
\ee
The right hand side of \eqref{eq:AlmostSplit-1} works out to be 
\be\label{eq:AlmostSplit-2} 
\begin{split} 
 \sum_{\vec \ell, \vec s \in \IZ_+^n} c(\vec \ell) c(\vec s)  
&
 \prod_x \left(  \sum_{d_x=0}^\infty (d_x \varepsilon_x^\vee)^{\ell_x} 
\frac{\lambda_x^{d_x}}{d_x!}\left( \frac{\theta_x^*}{\theta_x}\right)^{d_x}\left( d_x \frac{\varepsilon_x}{\theta_x}\right)^{s_x}\right) \\ 
\end{split} 
\ee 
So the desired splitting will only hold in the case that $\theta$ is real. 
In that case we  indeed have 
\be 
\bar\CA = \Phi_\theta \Phi_\theta^\dagger  ~ . 
\ee
Thus the existence of a splitting is not universally true, and can put 
constraints on the defining data of the seed TQFT.

\subsubsection{Relation To Coherent States In A Heisenberg Representation}\label{subsubsec:RelationCoherent}

We now can make contact with the literature on ``baby universe creation operators'' 
and wormholes  \cite{Balasubramanian:2020jhl,Gardiner:2020vjp,Marolf:2020xie}.
We take $b_\alpha >0$ so that the Hilbert space structure on (a suitable completion of) $S^\bullet \CC$
can be identified with  the standard Fock space representation of the Heisenberg algebra
associated to $\CC$.  For a single oscillator  $[a,a^\dagger]=1$, and $a \vert 0 \rangle =0$, and $\langle 0 \vert 0 \rangle=1$.
We can denote the unit norm vector 
$ \frac{1}{\sqrt{d!}} (a^{\dagger})^d \vert 0 \rangle$ simply by $ \vert d \rangle$ so that
\be
\langle d_1 \vert d_2 \rangle = \delta_{d_1, d_2} ~ .
\ee
This can trivially be extended to a collection of oscillators $a_x, a_x^\dagger$ labeled by $x\in {\rm Spec}(\CC)$. 
To make contact with standard physical notation
we identify:
\be
\vert d_1, \dots, d_s \rangle = \prod_{x=1}^n \left(\frac{\varepsilon_x}{\sqrt{\theta_x}}\right)^{d_x}
\ee

We now define the ``coherent vector'' (not to be confused with the Hartle-Hawking vector $\Psi_{HH}$ defined above!)
\be
\Psi_\theta = \sum_{\vec d \in \IZ_+^n}^\infty 
\prod_x \sqrt{\frac{\lambda_x^{d_x}}{d_x!} } \left(  \frac{\varepsilon_x}{\sqrt{\theta_x}} \right)^{\otimes d_x} \in S^\bullet \CC
\ee
In standard physics notation $\Psi_{\theta}$ corresponds to the coherent state: 
\be
\vert \lambda \rangle: = \exp\left( \sum_x \sqrt{\lambda_x} a_x^\dagger \right) \vert 0 \rangle ~ . 
\ee
We define a ``universe annihilating operator'' 
\be 
Z_{\rm an}: S^\bullet \CC \to \CC^\vee \otimes S^\bullet \CC 
\ee
by 
\be 
Z_{\rm an}: \vert d_1,\dots, d_n \rangle \mapsto 
\left(\sum_x d_x \varepsilon_x^\vee\right) \otimes  \vert d_1,\dots, d_n \rangle 
\ee
Let $\pi: \CC^{\otimes n} \to S^n \CC$ be the projection. We denote
\be
(\pi \otimes Id) \circ ( Id^{\otimes n}\otimes Z_{\rm an} ) \circ  ( Id^{\otimes n-1}\otimes Z_{\rm an} )\circ  \cdots \circ 
(1\otimes Z_{\rm an})\circ Z_{\rm an}
\ee
simply by $Z_{\rm an}^n$, so that 
\be 
Z_{\rm an}^n: S^\bullet \CC \to S^n\CC^\vee \otimes S^\bullet \CC 
\ee
Then we can say 
\be 
\Phi_{\theta}= \frac{1}{1-Z_{\rm an}} \Psi_{\theta}
\ee

Similarly 
\be 
\Phi_{\theta}^\dagger = \frac{1}{1- Z_{\rm cr} } \Psi_{\theta}^\vee
\ee
where 
\be 
Z_{\rm cr}: S^\bullet \CC^\vee \to S^\bullet \CC^\vee \otimes \CC 
\ee
is defined by 
\be 
Z_{\rm cr}: \langle d_1, \dots, d_n \vert \to \langle d_1, \dots, d_n
\vert \otimes \left( \sum_x d_x \frac{\varepsilon_x}{\theta_x} \right)
\ee
so now we can write 
\be\label{eq:TotalAmp-QMIP}
\bar \CA = \langle \bar\lambda \vert \frac{1}{1-Z_{\rm cr}} \frac{1}{1-Z_{\rm an}}\vert \lambda \rangle ~ . 
\ee

\textbf{Remarks}:

\begin{enumerate}

\item Note that $Z_{\rm an}$ and $Z_{\rm cr}$ are not really independent operators. 
We can write 
\be 
Z_{\rm an} \in S^\bullet \CC^\vee \otimes \CC^\vee \otimes S^\bullet \CC 
\ee
as 
\be 
Z_{\rm an} = \sum_d d \langle d \vert \otimes \varepsilon^\vee \otimes \vert d \rangle 
\ee
where
\be 
Z_{\rm cr} \in S^\bullet \CC \otimes S^\bullet \CC^\vee \otimes \CC 
\ee
is just 
\be 
Z_{\rm cr} = \sum_d \vert d \rangle \otimes \langle d \vert \otimes \frac{\varepsilon}{\theta} 
\ee
so that $Z_{\rm cr} = Z_{\rm an}^\dagger$. 

\item One of the primary points of the paper of Marolf and Maxfield \cite{Marolf:2020xie} 
is that their model supports the idea that holography fundamentally involves using 
expectation values in an \underline{ensemble} of field theories. Many of the formulae 
of this paper, and particularly the splitting formula that we emphasize, can be viewed 
as suggesting an alternative interpretation. The splitting space $\CW$ is the inner product space 
(or Hilbert space, if suitably completed) of a ``holographically dual'' quantum system. 
One way to read equation \eqref{eq:TotalAmp-QMIP} is that it states that expectation 
values   in the holographically dual quantum system with ``Hilbert space'' $\CW$ 
produce amplitudes of the quantum gravitational system. This interpretation 
resonates with ideas in \cite{Moore:2017wlq}. Note that this version of holography 
is somewhat more abstract than traditional examples since the ``dual quantum system'' 
is not associated with any asymptotic region of spacetime. It would be interesting to explore 
this alternative interpretation of summed amplitudes further. 

\end{enumerate}

\subsection{Non-Semisimple Theories}\label{subsec:NonSemiSimple}

In this sub-section we show that the splitting formula exists in two 
examples of non-semisimple theories. In \ref{subsubsec:GeneralNonSemiSimple} 
we discuss briefly what can be said about the general non-semisimple case.

\subsubsection{$\IC\IP^1$}

When  $\CC$ is two-dimensional and not semisimple.
We can think of this as the Frobenius structure on the cohomology of
$\IC\IP^1$ or the LG model with $W= X^3/3$.

We can choose a basis where  $e_1$ is the identity and $e_2$   is nilpotent so
\be
e_1^2 = e_1 \qquad  e_1 e_2 = e_2 \qquad e_2^2 = 0
\ee
  The trace is the integral so
\be
\begin{split}
\theta(e_1) & = 0 \\
\theta(e_2) & = \theta \\
\end{split}
\ee
where $\theta \in \kappa^*$ and $\kappa$ is the ground field.
It is natural to rescale $e_2$ so that $\theta = 1$.
Then the duality $\CC^\vee \cong \CC$ defined by the
nondegenerate pairing is:
\be
\begin{split}
b_\vee(e_1^\vee) = e_2 \qquad & \qquad b^\vee(e_1) = e_2^\vee \\
b_\vee(e_2^\vee)= e_1\qquad & \qquad b^\vee(e_2) = e_1^\vee \\
\end{split}
\ee
The $U$-shaped cylinder with 2 outgoing circles  mapped by $\CZ$ to 
\be
1 \in \kappa \mapsto e_1 \otimes e_2 + e_2 \otimes e_1
\ee
The comultiplication is:
\be
\begin{split}
\Delta(e_1) &  =  e_1 \otimes e_2 + e_2 \otimes e_1 \\
\Delta(e_2)& = e_2 \otimes e_2 \\
\end{split}
\ee

Then the characteristic element is $\fh=2e_2$ so the handle-adding operation is
multiplication of any output or input by
\be
\frac{1}{e_1-\fh} = e_1 +\fh = e_1 + 2 e_2
\ee
The sum over all connected surfaces  $\theta(1/1-\fh) = 2$.   This plays the role of $\lambda$.
So
\be
\bar\CA(0,0) = \exp[2]
\ee
We write $\exp$ and not the letter $e$ for the transcendental number $e=2.71828... $
in order to avoid confusion with our notation for vectors.

It is most straightforward to evaluate the Hartle-Hawking vector.
Define the symmetrization operation
\be
Sym^n (v_1 \otimes \cdots \otimes v_n) := \frac{1}{n!} \sum_{\sigma \in S_n} v_{\sigma(1)} \otimes \cdots \otimes v_{\sigma(n)}
\ee
Then the connected  bordism from $\emptyset \to (S^1)^n$ with no handles is
\be
\begin{split}
1 \in \kappa &  \mapsto e_1 \otimes e_2 \otimes e_2 \otimes \cdots \otimes e_2 + e_2 \otimes e_1 \otimes e_2 \otimes \cdots \otimes e_2
+ \cdots + e_2 \otimes e_2 \otimes e_2 \otimes \cdots \otimes e_2 \otimes e_1 \\
& =  n Sym^n (e_2^{\otimes (n-1)} \otimes e_1 )\\
&  = n e_2^{n-1} e_1\\
\end{split}
\ee
where the last equality is understood to be written in the symmetric algebra.
When we add handles we can, for example, add all the handles on the first of the outgoing
tubes so we multiply by  $(e_1 +2 e_2)\otimes 1^{\otimes (n-1)}$. For $j$ outgoing circles
this gives:
\be
x_j=  j e_1 e_2^{j-1} + 2 e_2^j  =  (2 + e_1 \frac{\p}{\p e_2}) e_2^j
\ee
again understood to be in the symmetric algebra.

So the component of the  HH state with $n$ outgoing boundaries is:
\be
\Psi_{HH} = \bar\CA(0,0)  B_n(x_1,\dots, x_n)
\ee
where $x_j$ are considered in the symmetric algebra $S^{\bullet} \CC$.
The first few terms are:
\be
\begin{split}
\bar\CA(0,0)^{-1} \Psi_{HH} & = 1 \\
& + (e_1 + 2 e_2) \\
& + (e_1^2 + 6 e_1 e_2 + 6 e_2^2)\\
& + (e_1^3 + 12 e_1^2 e_2 + 33 e_1 e_2^2 + 22 e_2^3) \\
& + \cdots \\
\end{split}
\ee

Using equation \eqref{eq:Bell-Def1} one derives  the $n$-th order term for $\Psi_{HH}$:
\be
 \left( \frac{\p}{\p t} \right)^n \vert_0  \left\{ \exp[ t e_1 e^{e_2 t} ] \cdot \exp[2 e^{e_2 t}] \right\}
\ee
which works out to
\be
 \sum_{d_2=0}^n \left(\sum_{d_1=0}^\infty \frac{2^{d_1}}{d_1!} \sum_{k=d_2}^n \frac{n!}{(n-k)! (k-d_2)!} d_1^{n-k} d_2^{k-d_2} \right) e_1^{d_2} e_2^{n-d_2}
\ee

Now it is not hard to show that
\be
\sum_{k=y}^n {n \choose k}{k \choose y}  x^{n-k} y^{k-y} = {n \choose y} (x+y)^{n-y}
\ee
and therefore the  full Hartle-Hawking vector is
\be
\Psi_{HH} =\sum_{a\geq0, b\geq 0 }  \sum_{d=0}^\infty \frac{2^d}{d!} (d+a)^b {a+b \choose b}  e_1^a e_2^b
\ee

 Using the rule \eqref{eq:bvee-rule} we get the full amplitude from arbitrary
ingoing to arbitrary outgoing states by
\be
\bar \CA = \sum_{m=0}^\infty (b^\vee)^m \Psi_{HH}
\ee
The result is, 
\be\label{eq:NewCP1-sum}
\bar\CA =   \sum_{d=0}^\infty \frac{2^d}{d!}\sum_{a_1, b_1, a_2, b_2\geq 0 } (d+a_1 + a_2)^{b_1 +b_2}
{a_1 + b_1 \choose a_1} {a_2 + b_2 \choose a_2} (e_2^\vee)^{a_1} (e_1^\vee)^{b_1} e_1^{a_2} e_2^{b_2}
\ee

We can now describe a splitting formula in the sense of \eqref{eq:SplittingFormula}. We wish
to write the amplitude in the form
\be\label{eq:CP1-FactoredForm}
\sum_{a_1, b_1} \sum_{a_2, b_2}  (e_2^\vee)^{a_1} (e_1^\vee)^{b_1} \left( \sum_x N(a_1, b_1;x) M(a_2, b_2; x)\right) e_1^{a_2} e_2^{b_2}
\ee
where the sum on $x$ is independent of $a_1,b_1,a_2,b_2 $. We will achieve this
but there  is no claim that this is the unique way of factorizing the amplitudes and there might
well be a physically more transparent splitting presentation of \eqref{eq:CP1-FactoredForm}.

Expanding the factors  $(d+a_1 + a_2)^{b_1} $ and $(d+a_1 + a_2)^{b_2} $
using the binomial theorem, and after a surprising reshuffling of combinatoric factors, 
we find:  
\be
\begin{split}
\sum_{a_1, b_1} \sum_{a_2, b_2} & (e_2^\vee)^{a_1} (e_1^\vee)^{b_1}\\
  \sum_{d, k_1, k_2}  {a_1 + b_1 \choose a_1}
\frac{2^d}{d!} {b_1 \choose k_1} (d+a_1)^{b_1 - k_1} a_1^{k_2}
&  {a_2 + b_2 \choose a_2}  {b_2 \choose k_2} (d+a_2)^{b_2 - k_2} a_2^{k_1} \\
  &  e_1^{a_2} e_2^{b_2}\\
  \end{split}
\ee

We now introduce an infinite-dimensional \underline{inner product space}, $\CW$,
equipped with a basis $\{ u_{d,k_1,d_2} \}$ with $d,k_1,k_2 \in \IZ_{\geq 0}$. The
 inner product is defined to be
\be
\langle  u_{d,k_1, k_2},  u_{d',k_1', k_2'} \rangle = \delta_{d,d'} \delta_{k_1, k_2'} \delta_{k_2,k_1'}
\ee
Note that with this inner product we have
\be
(u_{d,k_1,k_2})^\dagger = u_{d,k_2,k_1}^\vee
\ee
Now define
\be
\Phi: S^\bullet \CC \rightarrow \CW
\ee
by
\be
\Phi =  \sum_{a_1, b_1} \sum_{d,k_1,k_2}
(e_2^\vee)^{a_1} (e_1^\vee)^{b_1}
 \left\{ \sqrt{ \frac{2^d}{d!}} {a_1+b_1 \choose a_1}  {b_1 \choose k_1} (d+a_1)^{b_1 - k_1} a_1^{k_2} \right\}
  u_{d,k_1,k_2}
\ee
Then it is easy to check
\be
\bar\CA = \Phi \Phi^\dagger ~ . 
\ee

\subsubsection{$\IC \IP^2$}
We consider another example, that of the Frobenius algebra corresponding to the cohomology ring of $\IC\IP^2$. A basis for this algebra (and the product operation) is the following,
\begin{equation}
    \begin{split}
        &e_1^2=e_1 \quad e_2^2=e_3\quad e_3^2=0 \\
        &e_1e_2=e_2 \quad e_1e_3=e_3 \quad e_2e_3=0
    \end{split}
\end{equation}
and we see that $e_1$ plays the role of the unit. The counit operation, after the usual rescaling, is
\begin{equation}
    \theta(e_1)=\theta(e_2)=0 \quad \theta(e_3)=1
\end{equation}
The comultiplication acts on the above basis in the following manner,
\begin{equation}
    \begin{split}
        &\Delta(e_1)=e_1\otimes e_3+e_2\otimes e_2+e_3\otimes e_1 \\
        &\Delta(e_2)=e_2\otimes e_3+e_3\otimes e_2 \\
        &\Delta(e_3)=e_3\otimes e_3
    \end{split}
\end{equation}
We can now find the characteristic element (defined as $m(\Delta(e_1))$, where $m$ is the product operation),
\begin{equation}
    \mathfrak{h}=3e_3
\end{equation}
which squares to zero. Also, we can confirm the dimension of the algebra from the equation $\theta(\mathfrak{h})=3$, which is the TQFT amplitude on the torus. Now, summing over all powers of the characteristic element gives us,
\begin{equation}
    H=\sum_{g=0}^{\infty}\mathfrak{h}^g=e_1+\mathfrak{h}=e_1+3e_3
\end{equation}
The sum over all connected bordisms is then,
\begin{equation}
    \theta\bigg(\sum_{g=0}^{\infty}\mathfrak{h}^g\bigg)=\theta(H)=3
\end{equation}
which is the analogue of $\lambda$. The sum over all bordisms without boundary then becomes,
\begin{equation}
    \mathcal{A}(0,0)=\text{exp}[3]
\end{equation}
We can now evaluate amplitudes corresponding to sums over bordisms with boundaries. For this, note that 
\begin{equation}
    \theta(e_1H)=3, \quad \theta(e_2H)=0, \quad \theta(e_3H)=1
\end{equation}
This tells us how to set up the combinatorics for general amplitudes.
\begin{enumerate}
    \item Let us begin by considering amplitudes with only ingoing boundaries. For outgoing boundaries, we will simply need to apply the duality maps. We consider a case where we have $n$ ingoing boundary circles labeled $e_1$, $m$ circles labeled $e_3$ and $(2r)$ circles labeled $e_2$, where $m,n,r\in\{0,1,2,\cdots\}$.
    \item Note that we need an even $(2r)$ connected boundary components labeled with the state $e_2$ since for an odd number of such boundaries, the amplitude vanishes. We start by pairing these components into $r$ pairs. This can be done in 
    \begin{equation}
        \frac{(2r)!}{2^rr!} \text{ ways}
    \end{equation}
    \item For each positive integer $k\in\{0,1,2,\cdots,n\}$, we now choose $(n-k)$ of the boundary circles labeled $e_1$ and connect them among themselves. This can be done in
    \begin{equation}
        {n\choose k} \text{ ways}
    \end{equation}
    and leads to a Bell polynomial
    \begin{equation}
        B_{n-k}(3)\text{exp}[3]=\sum_{d=0}^{\infty}\frac{3^d}{d!}d^{n-k}
    \end{equation}
    \item The rest of the $k$ circles labeled $e_1$ now need to be collected into bordisms connected to either the $m$ circles labeled $e_3$ or the $r$ pairs of circles labeled $e_2\otimes e_2$. This can be done in $(m+r)^k$ ways.
    \item Finally, the ingoing boundary condition belongs to $S^{\bullet}(\mathcal{C})$, so we need to symmetrize the boundary condition, leading to a factor of
    \begin{equation}
        \frac{(n+2r+m)!}{n!(2r)!m!}
    \end{equation}
\end{enumerate}
The complete contribution then becomes
\begin{equation}
    \begin{split}
        &\mathcal{A}(m+n+2r,0)(\text{Sym}(e_1^{\otimes n}\otimes e_2^{\otimes 2r}\otimes e_3^{\otimes m}))=\sum_{d=0}^{\infty}\frac{3^d}{d!}\sum_{k=0}^n {n\choose k}d^{n-k}(m+r)^k\frac{(2r)!}{2^rr!}\frac{(n+2r+m)!}{n!(2r)!m!}\\
        &=\sum_{d=0}^{\infty}\frac{3^d}{d!}(d+m+r)^n\frac{(2r)!}{2^rr!}\frac{(n+2r+m)!}{n!(2r)!m!} 
    \end{split}
\end{equation}
We can now similarly work with amplitudes corresponding to bordisms with only nonempty outgoing boundary to write the Hartle-Hawking vector,
\begin{equation}
    \Psi_{HH}=\mathcal{A}(0,0)^{-1}\sum_{m,n,r=0}^{\infty}\sum_{d=0}^{\infty}\frac{3^d}{d!}(d+m+r)^n\frac{(2r)!}{2^rr!}\frac{(n+2r+m)!}{n!(2r)!m!}\text{Sym}((e_1)^{\otimes m}\otimes (e_2)^{\otimes 2r}\otimes (e_3)^{\otimes n})
\end{equation}
The general amplitude with an arbitrary number of ingoing and outgoing amplitudes then becomes,
\begin{equation}
    \begin{split}
        &\sum_{m_1,n_1,r_1\geq 0}\sum_{m_2,n_2,r_2\geq 0}\sum_{d=0}^{\infty}\frac{3^d}{d!}(d+m_1+m_2+r_1+r_2)^{n_1+n_2}\frac{(2r_1+2r_2)!}{2^{r_1+r_2}(r_1+r_2)!} \\
        &\times \frac{(n_1+(2r_1)+m_1)!}{n_1!(2r_1)!m_1!}\frac{(n_2+(2r_2)+m_2)!}{n_2!(2r_2)!m_2!}
        (e_1^{\vee})^{n_1}(e_2^{\vee})^{2r_1}(e_3^{\vee})^{m_1}e_1^{m_2}e_2^{2r_2}e_3^{n_2} 
    \end{split}
\end{equation}
We see that there is a new factor,
\begin{equation}
    \frac{(2r_1+2r_2)!}{2^{r_1+r_2}(r_1+r_2)!}
\end{equation}
that will hinder factorization, but this is precisely the factor we encountered in the case of 1d TQFTs. We can therefore proceed as before to obtain,
\begin{equation}
    \begin{split}
        &\sum_{m_1,n_1,r_1\geq 0}\sum_{m_2,n_2,r_2\geq 0}\sum_{d=0}^{\infty}\frac{3^d}{d!}(d+m_2+r_2+m_1+r_1)^{n_2}(d+m_2+r_2+m_1+r_1)^{n_1}\frac{(2r_1+2r_2)!}{2^{r_1+r_2}(r_1+r_2)!} \\
        &\times \frac{(n_1+(2r_1)+m_1)!}{n_1!(2r_1)!m_1!}\frac{(n_2+(2r_2)+m_2)!}{n_2!(2r_2)!m_2!}
        (e_1^{\vee})^{n_1}(e_2^{\vee})^{2r_1}(e_3^{\vee})^{m_1}e_1^{m_2}e_2^{2r_2}e_3^{n_2} 
    \end{split}
\end{equation}
We use the binomial expansion as before,
\begin{equation}
    \begin{split}
        &=\sum_{m_1,n_1,r_1\geq 0}\sum_{m_2,n_2,r_2\geq 0}\sum_{d=0}^{\infty}\frac{3^d}{d!}\sum_{k_2=0}^{n_2}{n_2\choose k_2}(d+m_2+r_2)^{n_2-k_2}(m_1+r_1)^{k_2}\sum_{k_1=0}^{n_1}{n_1\choose k_1}(d+m_1+r_1)^{n_1-k_1} \\
        &\times(m_2+r_2)^{k_1} \frac{(2r_1+2r_2)!}{2^{r_1+r_2}(r_1+r_2)!}\frac{(n_1+(2r_1)+m_1)!}{n_1!(2r_1)!m_1!}\frac{(n_2+(2r_2)+m_2)!}{n_2!(2r_2)!m_2!}
        (e_1^{\vee})^{n_1}(e_2^{\vee})^{2r_1}(e_3^{\vee})^{m_1}e_1^{m_2}e_2^{2r_2}e_3^{n_2} 
    \end{split}
\end{equation}
We now split the extra factor as in the 1d theory,
\begin{equation}
    \begin{split}
        &=\sum_{m_2,n_2,r_2\geq 0}\sum_{m_1,n_1,r_1\geq 0}\sum_{d=0}^{\infty}\sum_{k_1,k_2}\frac{3^d}{d!}\int_{-\infty}^{\infty}\frac{dh}{\sqrt{2\pi}}e^{-\frac{1}{2}h^2} \\
        &\times \bigg[{n_2\choose k_2}(d+m_2+r_2)^{n_2-k_2}(m_2+r_2)^{k_1}h^{2r_2}\frac{(n_2+(2r_2)+m_2)!}{n_2!(2r_2)!m_2!}(e_1)^{m_2}(e_2)^{2r_2}(e_3)^{n_2}\bigg] \\
        &\times \bigg[{n_1\choose k_1}(d+m_1+r_1)^{n_1-k_1}(m_1+r_1)^{k_2}h^{2r_1}\frac{(n_1+(2r_1)+m_1)!}{n_1!(2r_1)!m_1!}(e_1^{\vee})^{n_1}(e_2^{\vee})^{2r_1}(e_3^{\vee})^{m_1}\bigg]
    \end{split}
\end{equation}
and we can once again introduce an auxiliary Hilbert space $\mathcal{W}$ spanned by elements of the form
\begin{equation}
    u_{d,k_1,k_2,h}
\end{equation}
to write the amplitudes in split form,
\begin{equation}
    \mathcal{A}=\Phi\Phi^{\vee}, \quad \Phi:S^{\bullet}(\mathcal{C})\to\mathcal{W}
\end{equation}
where,
\begin{equation}
    \begin{split}
        &\Phi=\sum_{m_2,n_2,r_2}\sum_{d,k_1,k_2}\sqrt{\frac{3^d}{d!}}{n_2\choose k_2}\frac{(n_2+2r_2+m_2)!}{n_2!(2r_2)!m_2!}\int_{-\infty}^{\infty}\frac{dh}{(2\pi)^{1/4}}e^{-\frac{1}{4}h^2} \\
        &(d+m_2+r_2)^{n_2-k_2}(m_2+r_2)^{k_1}h^{2r_2}(e_1)^{m_2}(e_2)^{2r_2}(e_3)^{n_2}u_{d,k_1,k_2,h}
    \end{split}
\end{equation}
In the $\mathbb{CP}^N$ cohomology ring TQFTs for higher $N$, we will have more such factors of the form $p!/q!$, and a general way to factorize such terms involves using the integral forms of the Gamma function and the reciprocal Gamma function.

\subsubsection{Some Remarks On The General Non-Semisimple Case}\label{subsubsec:GeneralNonSemiSimple} 

It was noted in \cite{Sawin:1995rh}   that since a 2d closed TQFT is 
completely determined by a commutative Frobenius algebra 
we can decompose the theory according to an orthogonal 
decomposition of Frobenius algebras. Therefore we can 
decompose it into a semisimple and non-semisimple part. 
The non-semisimple subalgebra will have nilpotent elements. 
In general the non-semisimple subalgebra can be decomposed as a 
sum of irreducible algebras $A$. For the irreducible algebras 
there will be a filtration by  ideals
\be 
A = N_m \supset N_{m-1} \supset \cdots \supset N_0 
\ee
where $N_0$ is the so-called \emph{socle}, the 
space of all elements $x\in A$ with the property 
that $a x=0$ for all $a\in A$ which are nilpotent. 
In the Frobenius case the socle is one-dimensional, 
and there is a unique element $s\in N_0$ such that 
$\theta(s)=1$. One can show that there is a basis 
$\{a_1, \dots, a_n\}$ for $A$ and a dual basis 
$\{b_1, \dots, b_n \}$ for $A$, so $\theta(a_i b_j) = \delta_{ij}$, 
and such that   $a_i b_i = s$. A good example of this structure 
is provided by the
cohomology of an arbitrary manifold, 
\footnote{Assume for simplicity that the odd dimensional cohomology vanishes. If not then the Frobenius algebra is super-commutative and many signs need to be accounted for.}
$A= H^*(M;\kappa)$. Then if $\dim M = m$ we can 
take $N_k$ to be the subspace of forms of degree $\geq m-k$. 
Then we take $a_1 =1, \dots, a_n = s$ and $b_1 = s, \dots, b_n = 1$ 
and $\{ a_i\} $ and $\{ b_j\} $ are Poincar\'e dual bases. 
One can show that the comultiplication is 
\be 
\Delta(x) = \sum_i  x a_i \otimes b_i = \sum_i a_i \otimes b_i x 
\ee
and $\fh = (\dim A) s$ so $\CA(0,0) = \exp[\dim A]$. To go further 
one needs to make use of the ring structure of $A$. For example, 
the genus zero connected bordism from  $\emptyset$ to the disjoint 
union of three circles maps $1\in \kappa$ to 
\be 
\sum_{i,j}  a_i a_j \otimes b_j \otimes b_i  ~ . 
\ee
Clearly to go further and produce an explicit formula for the HH vector 
we need to make use of the ring structure  $a_i a_j = c_{ij}^k a_k$. 
We leave further analysis for another occasion.

\section{$d=2$ Open-Closed TQFT}

We now come to the case of open-closed TQFT. We will only
discuss the semisimple case, so this is the same thing as
fully extended 2d TQFT.

The open-closed theory is characterized by a $\kappa$-linear category
of boundary conditions $\fB$. As shown in \cite{Moore:2006dw}, in the
semisimple case the general structure is the following.
We can decompose the closed string space $\CC$ into idempotents
as in equation \eqref{eq:Decomp-x} above. Then we must choose a square-root
$\mu_x$ of each of the ``dilaton values'' $\theta_x = \mu_x^2$.
\footnote{As explained in \cite{Moore:2006dw} the choice of square root $\mu$ 
is a primitive version
of the ``$B$-field'' of string theory.}
and the general boundary condition $a \in Obj(\fB)$ is a
finite-dimensional vector bundle $W \to {\rm Spec}(\CC)$.
Concretely, this is a vector space $W_x$ attached to each $\varepsilon_x$.

\subsection{Very Simplest Case}

Let us take the semisimple case with $\CC = \kappa \cdot \varepsilon$ and $\varepsilon^2=\varepsilon$ and $\theta(\varepsilon) = \mu^2$
where we have a chosen $\mu$. Let us focus on a single boundary condition $a$ with $\CO_{aa} = Mat_{N_a\times N_a}(\kappa)$.

In this case we have the open to closed map on $T\in \CO_{aa}$
\be
\iota^a(T) =  \frac{\Tr(T)}{\mu } \varepsilon
\ee
and the open trace:
\be
\theta_a(T) = \mu \Tr(T) ~ . 
\ee
A natural basis for $\CO_{aa}$ is $e_{ij}$ and then
\be
\begin{split}
b^\vee(e_{ij}) & = \mu e_{ji}^\vee \\
b_{\vee}(e_{ij}^\vee) & = \frac{1}{\mu} e_{ji} ~ . \\
\end{split}
\ee

Note that in the sum over bordisms just as the closed-string amplitudes factor descend to endomorphisms of $S^{\bullet}\CC$, similarly the
open-string amplitudes descend to an endomorphism of $S^\bullet\CO_{aa}$. We will now give an explicit formula for this endomorphism in terms of the
defining data of the open-closed theory and exhibit the basic splitting phenomenon of equation \eqref{eq:SplittingFormula}.

We first derive a formula for the Hartle-Hawking covector.
The amplitude associated with the sum of bordisms from
one ingoing open interval to the emptyset can be thought of as the composition
of maps   $T \in \CO_{aa}\to \CC \to \kappa$ and is given explicitly by
\be
T\to \frac{\Tr(T)}{\mu } \varepsilon \to \frac{\Tr(T)}{\mu }e^\lambda B_1(\lambda)
\ee
where $T\in \CO_{aa}$.

Now consider the sum of bordisms from two ingoing open intervals to the emptyset.
With two ingoing open intervals we must contract the indices of $T_1 \otimes T_2  \in \CO_{aa} \otimes \CO_{aa}$.
This can be done in two ways. The resulting state can then be mapped to a closed string state, and then
we can map the resulting closed string state to $\kappa$ in the two ways counted by the Bell polynomials.
\paragraph{} 
\noindent\rule[0.5ex]{\linewidth}{1pt}
\\
\usetikzlibrary{intersections, pgfplots.fillbetween}
\usetikzlibrary{decorations.markings}
\begin{tikzpicture}
   [decoration={markings, 
    mark= at position 0.5cm with {\arrow{stealth}}}
]
   \draw [postaction={decorate}] (0,0) -- (0,1);
   \draw [postaction={decorate}] (0,2) -- (0,3);
   \draw [dash pattern=on 1pt off 2\pgflinewidth] (0,0) -- (2,0);
   \draw [dash pattern=on 1pt off 2\pgflinewidth] (0,1) -- (2,1);
   \coordinate (O) at (2,0);
   \coordinate (A) at (2,1);
   \coordinate (B) at (2,2);
   \coordinate (C) at (2,3);
   \draw [dash pattern=on 1pt off 2\pgflinewidth] (0,2) -- (2,2);
   \draw [dash pattern=on 1pt off 2\pgflinewidth] (0,3) -- (2,3);
   \draw [fill=gray] (O) to [bend right=20] (A) to [bend right=50] (B) to [bend right=20] (C) to [bend left=50] (O);
   \draw [postaction={decorate}] (6,0) -- (6,1);
   \draw [postaction={decorate}] (6,2) -- (6,3);
   \draw [dash pattern=on 1pt off 2\pgflinewidth] (6,0) -- (8,0);
   \draw [dash pattern=on 1pt off 2\pgflinewidth] (6,1) -- (8,1);
   \draw [dash pattern=on 1pt off 2\pgflinewidth] (6,2) -- (8,2);
   \draw [dash pattern=on 1pt off 2\pgflinewidth] (6,3) -- (8,3);
   \coordinate (D) at (8,0);
   \coordinate (E) at (8,1);
   \coordinate (F) at (8,2);
   \coordinate (G) at (8,3);
   \draw [dash pattern=on 1pt off 2\pgflinewidth, name path=P1](D) to [bend right=20] (E);
   \draw [dash pattern=on 1pt off 2\pgflinewidth, name path=Q1](F) to [bend right=20] (G);
   \draw [black, name path=P2] plot [smooth, tension=1] coordinates { (8,0) (8.5,0.2) (8.5,0.8) (8,1)};
   \draw [black, name path=Q2] plot [smooth, tension=1] coordinates { (8,2) (8.5,2.2) (8.5,2.8) (8,3)};
   \tikzfillbetween[of=P1 and P2]{gray};
   \tikzfillbetween[of=Q1 and Q2]{gray};
   \draw [postaction={decorate}] (12,0) -- (12,1);
   \draw [postaction={decorate}] (12,2) -- (12,3);
   \draw [dash pattern=on 1pt off 2\pgflinewidth] (12,0) -- (14,0);
   \draw [dash pattern=on 1pt off 2\pgflinewidth] (12,1) -- (14,1);
   \draw [dash pattern=on 1pt off 2\pgflinewidth] (12,2) -- (14,2);
   \draw [dash pattern=on 1pt off 2\pgflinewidth] (12,3) -- (14,3);
   \coordinate (H) at (14,0);
   \coordinate (J) at (14,1);
   \coordinate (K) at (14,2);
   \coordinate (L) at (14,3);
   \draw [dash pattern=on 1pt off 2\pgflinewidth](J) to [bend right=20] (K);
   \draw [dash pattern=on 1pt off 2\pgflinewidth, name path=N1](H) to [bend right=20] (L);
   \draw [black, name path=N2] plot [smooth, tension=1] coordinates { (14,0) (15.5,0.2) (15.5,2.8) (14,3)};
   \tikzfillbetween[of=N1 and N2]{gray};
\end{tikzpicture}

\paragraph{} 

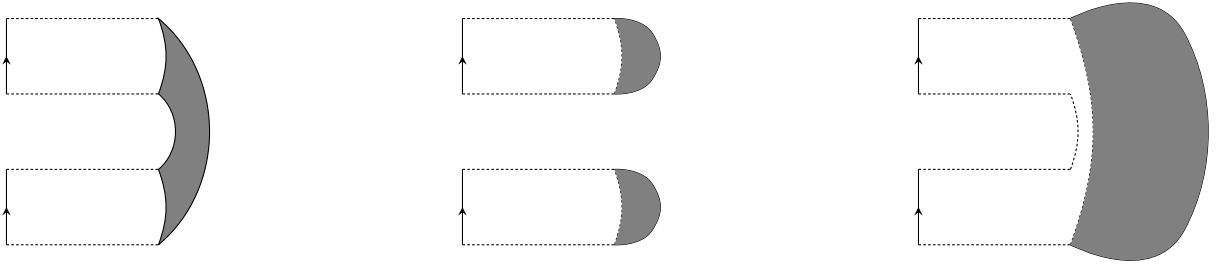
\textbf{Figure 3:} \emph{The three contributions to the amplitude corresponding to the sum over bordisms with two interval boundaries}
\\
\noindent\rule[0.5ex]{\linewidth}{1pt}

Thus, if we connect the dotted lines on the ends of the one-dimensional morphisms to themselves we get the
closed string state
\be
\frac{\Tr(T_1)}{\mu } \varepsilon \otimes  \frac{\Tr(T_2)}{\mu } \varepsilon
\ee
Then we map the two ingoing closed string states to nothing either via two connected components or via one
connected component. So we get
\be
\frac{\Tr(T_1)}{\mu }   \frac{\Tr(T_2)}{\mu } B_2(\lambda) e^\lambda ~ . 
\ee
Alternatively, we can connect the dotted lines together to get the closed string state
\be
  \frac{\Tr(T_1 T_2)}{\mu } \varepsilon
\ee
and then this is mapped to $\kappa$ to give
\be
\frac{\Tr(T_1 T_2)}{\mu } B_1(\lambda)e^{\lambda} ~ . 
\ee
We sum these two so at degree two the HH covector acts on 
$T_1 T_2 \in S^2 \CO_{aa}$ as follows
\be
\Psi_{HH}^\vee:  T_1 T_2 \rightarrow  e^\lambda \left(
\frac{\Tr(T_1)}{\mu }   \frac{\Tr(T_2)}{\mu } B_2(\lambda) + \frac{\Tr(T_1 T_2)}{\mu } B_1(\lambda) \right) ~ . 
\ee

The above reasoning generalizes. It is helpful at this point to recall the categorical viewpoint: 
The ``open string'' intervals, either ingoing or outgoing, are mapped by the field theory functor $\CZ$ to morphisms in a 1-category $\fB$, the category of boundary conditions. The interval 
is therefore oriented and the $k^{th}$ interval in general maps boundary condition $a_k$ to $b_k$ where 
$a_k$ and $b_k$ are objects in $\fB$. The two-dimensional surfaces are mapped by $\CZ$ to 2-morphisms in the target category $\fC$. 
\footnote{The 2-category is the 2-category of (Frobenius) algebras.  The objects (i.e. the $0$-morphisms) are Frobenius algebras, the $1$-morphisms are bimodules, and the $2$-morphisms are intertwiners.}
The image of a 2-dimensional bordism maps the images of the  ingoing  intervals to those of the outgoing intervals. 
As the two-dimensional bordism ``evolves'' the endpoints of each interval ``evolve.'' In figures 
3,4,5 below the evolution of these endpoints is illustrated as a dotted line
 carrying the appropriate boundary condition. 
In \cite{Moore:2006dw} such boundaries were called 
\emph{constrained boundaries} and we will continue to use that terminology in this paper. 
The dotted lines with the same boundary condition 
must be connected up so that an ``outgoing object'' (in the sense of a 1-morphism) is
connected to an ``ingoing object'' (in the sense of a 1-morphism). 
The relevant intervals might be in- or out-going in the 2-morphism.

\clearpage 
\noindent\rule[0.5ex]{\linewidth}{1pt}
\\
\usetikzlibrary{decorations.markings}
\begin{tikzpicture}
[decoration={markings, 
    mark= at position 0.5cm with {\arrow{stealth}}}
] 
\coordinate (A1) at (0,0);
\coordinate (A2) at (0,1);
\coordinate (B1) at (0,2);
\coordinate (B2) at (0,3);
\coordinate (C1) at (0,4);
\coordinate (C2) at (0,5);
\draw [postaction={decorate}] (A1)--(A2);
\draw [postaction={decorate}] (B1)--(B2);
\draw [postaction={decorate}] (C1)--(C2);
\begin{scope}[dash pattern=on 1pt off 2\pgflinewidth, very thick,-] 
  \draw (A1) to [dash pattern=on 1pt off 2\pgflinewidth] (4,0);
  \draw (4,1) to [dash pattern=on 1pt off 2\pgflinewidth] (A2);
  \draw (B1) to [dash pattern=on 1pt off 2\pgflinewidth] (4,2);
  \draw (4,3) to [dash pattern=on 1pt off 2\pgflinewidth] (B2);
  \draw (C1) to [dash pattern=on 1pt off 2\pgflinewidth] (4,4);
  \draw (4,5) to [dash pattern=on 1pt off 2\pgflinewidth] (C2);
\end{scope}
\draw [dash pattern=on 1pt off 2\pgflinewidth, fill=gray]
  (4,-1) to (4,6)
  (4,6) to [bend left=50] (4,-1);
\end{tikzpicture}
\\
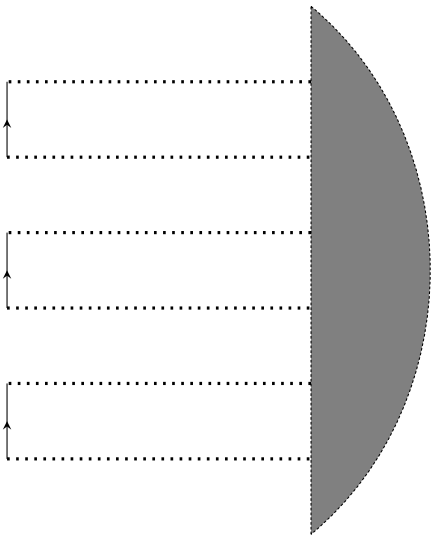
\textbf{Figure 4:} \emph{The ingoing and outgoing intervals have an orientation. They should be considered one-morphisms in a two-category 
between objects of that two-category, while the two-dimensional surface is a two-morphism between one-morphisms. The dashed lines inherit an orientation and carry a boundary condition associated with the end of the ingoing (or outgoing) interval. The dashed lines with the same boundary condition must be connected up, leading to symmetric group combinatorics when evaluating the amplitudes.}
\\
\noindent\rule[0.5ex]{\linewidth}{1pt}
\\
\paragraph{} 

Now consider the the case where all the ingoing intervals are morphisms of a single boundary 
condition $a\in Obj(\fB)$ to itself. 
With $n$ ingoing intervals there are $n$ outgoing (in the sense of the 1-morphism) 
dotted lines which must connect
up with one of the $n$ ingoing (in the sense of the 1-morphism) dotted lines. 
There is one diagram for each
element of the symmetric group $\sigma \in S_n$, telling which ingoing line is the destination
of a given outgoing line. Thus, for each element of the symmetric group, $\sigma \in S_n$ we associate
an outgoing closed string state:
\be
\otimes_{c\in cyc(\sigma)} \frac{ \Tr(T_c) }{\mu} \varepsilon
\ee
where $cyc(\sigma)$ is the set of cycles in the cycle decomposition of $\sigma$.
So $\sigma = \prod_{cyc(\sigma)} c$ where $c$ are disjoint cycles and
 for each cycle $c$ in that decomposition we can write  $c= (i_1, \dots, i_k)$. We
denote
\be
\Tr(T_c):= \Tr( T_{i_1} \cdots T_{i_k}) ~ . 
\ee

Suppose there are $r(\sigma)$ cycles in the cycle decomposition of $\sigma$. Now we can map the closed
strings to nothing using the usual Bell polynomial combinatorics.
Thus, the map
\be
\bar\CA^{\rm open}(n,0): \CO \otimes \cdots \otimes \CO \to \kappa
\ee
is given by
\be
T_1  \cdots   T_n  \mapsto \sum_{\sigma \in S_n}
  B_{r(\sigma)}(\lambda) e^{\lambda}\left( \prod_{c\in cycl(\sigma)} \mu^{-1} \Tr(T_c) \right)
\ee
where $T_1 \cdots T_n \in S^n\CO_{aa}$.  We will henceforth just write 
$r(\sigma)$ simply as   $r$ to lighten the notation.
Using the usual formula for the Bell polynomial we can  write:
\be\label{eq:Open-HHCV-deg-n}
T_1  \cdots   T_n  \mapsto \sum_{d=0}^\infty \frac{\lambda^d}{d!} \sum_{\sigma \in S_n}
\left( \frac{d}{\mu} \right)^r
(\prod_{c\in cycl(\sigma)} \Tr(T_c) ) ~ . 
\ee

Now, we can give a nice formula for the HH covector.

Consider first a fixed summand associated to $d$ in equation \eqref{eq:Open-HHCV-deg-n}.
We define the   functionals on $S^n\CO$:
\be\label{eq:varphi-n-def}
\varphi_{n,x}(T_1,\dots, T_n) := \sum_{\sigma \in S_n}  \prod_{c} ( x \Tr T_c)   ~ .
\ee
and we have $x=d/\mu$
for the $d^{th}$ term in \eqref{eq:Open-HHCV-deg-n}.
 Now define $\tilde \varphi_{n,x}(T)= \varphi_{n,x}(T,\dots, T)$
to be the restriction to the diagonal.
(Recall, again the remark of Appendix \ref{App:MultiLinear-Diagonal}.)
 We claim that
\be\label{eq:GenSum-varphi}
 \sum_{n\geq 0} \frac{\tilde \varphi_{n,x}(T)}{n!}  = \exp[-x \log\det(1-T)]
\ee
To prove this note that it suffices to prove it for the case $N_a=1$, so $\CO_{aa}=\kappa$,  since
the generic matrix is diagonalizable and the functional is continuous.
Setting $T=t\in \kappa$ equation \eqref{eq:varphi-n-def} simplifies to
\be
\tilde\varphi_{n,x}(t) = \sum_{P_n} \frac{n!}{\prod_{k=1}^n k^{\ell_k} \ell_k! } x^{\sum_k \ell_k} t^n
\ee
where $P_n$ is the set of partitions of $n$, so $1\cdot \ell_1 + 2\cdot \ell_2 + \cdots + n\cdot \ell_n = n$.
These partitions are in 1-1 correspondence with conjugacy classes in $S_n$. Now notice that
\be
\sum_{n=0} \frac{\tilde\varphi_{n,x}(t)}{n!} = \exp[ - x \log(1-t) ]
\ee
from which \eqref{eq:GenSum-varphi} follows.

We now observe that there is a matrix model representation
\be
\sum_{n\geq 0} \frac{\tilde\varphi_{n,x}(T)}{n!}= \int_{\CE_{N_a}}  [dH]  e^{- Tr V_x(H) + Tr(TH)}
\ee
where $\CE_{N_a}$ is the ensemble of $N_a\times N_a$ Hermitian matrices and $[dH]$ is the
standard Euclidean measure.
We can derive $V_x(H)$ from the inverse matrix Laplace transform. It is given by:
\be
e^{-V_x(H)} = \int_{\sqrt{-1} \CE_{N_a}}  [dS] e^{-x\Tr\log(1-S) - \Tr(SH) }
\ee
where $[dS]$ is the Euclidean measure divided by $(2\pi)^{N_a^2}$.  
Thus we obtain the matrix-model representation:
\be
 \varphi_{n,x}(T) = \int_{\CE_{N_a}}  [dH]  e^{- Tr V_x(H) } \prod_{i=1}^n \Tr(H T_i) ~ . 
\ee

Finally, including the sum over $d$ we have
\be
\Psi_{HH}^\vee(T_1, \dots, T_n) = \int_{\CE_{N_a}} [dH] e^{-U(H)} \prod_{i=1}^n \Tr(H T_i)
\ee
where
\footnote{The existence of the integral is tricky. It depends on the sign of $\mu$ and might need to be defined by analytic continuation, and it might be distribution-valued, as a function of $H$.} 
\be
\begin{split}
e^{- U(H)} & =  \int_{\sqrt{-1} \CE_{N_a}}  [dS] \exp \biggl[\lambda e^{- \frac{1}{\mu} \Tr\log(1-S)}  - \Tr(SH) \biggr]\\
& =  \int_{\sqrt{-1} \CE_{N_a}}  [dS] \exp \biggl[\frac{\lambda}{(\det(1-S))^{1/\mu} }  - \Tr(SH) \biggr] ~ . \\
\end{split}
\ee

Now, applying the $b_\vee$ operators we get
\be
\bar\CA = \sum_{S_i, S_o }     \prod_{a=1}^{n_i} e^\vee_{i_a, j_a} \prod_{b=1}^{n_o} \frac{e_{j_b i_b}}{\mu}
 \int_{\CE_{N_a}} [dH] e^{-U(H)}  \prod_{S_i} H_{j_a,i_a}  \prod_{S_o} H_{j_b, i_b}
\ee
Here $S_i$ is a collection $i_1, j_1, \dots i_{n_{\rm in}}, j_{n_{\rm in}}$ and similarly for $S_o$.
It follows rather directly that we have
\be\label{eq:Open-Amp-Split}
\bar\CA = \Phi_{a,\mu} \Phi_{a,\mu}^\dagger
\ee
where
\be
\Phi_{a,\mu}: S^\bullet \CO \to L^2(\CE_{N_a})^\vee
\ee
is given by
\be\label{eq:Open-Phi-a}
\Phi_{a,\mu} :=  \sum_{S_i }     \prod_{a=1}^{n_i} e^\vee_{i_a, j_a}
 \int_{\CE_{N_a}} [dH] e^{-\half U(H)}  \prod_{S_i} H_{j_a,i_a}  \langle H \vert
\ee
and, similarly to what we found in the 1d case, $\mu$ must be 
real. (In particular, $\theta_x>0$.) 

With the sesquilinear form such that
\be\label{eq:Open-SesquiForm}
(e^\vee_{ij})^\dagger = \mu^{-1} e_{ij}  \quad .
\ee
the sesquilinear structure on $\CO$ is nondegenerate 
and real, but not necessarily positive.

\textbf{Remarks}:

\begin{enumerate}

\item The above formula recovers nicely a formula of \cite{Gardiner:2020vjp} 
(subject to the remarks of section \ref{subsec:RelateMM} above) which, in our notation, 
is the HH covector evaluated on the exponentiated diagonal: 
\be\label{eq:GardMeg-formula}
\Psi_{HH}^\vee(e^T) = \exp[ \frac{\lambda}{(\det(1-T))^{1/\mu} }] ~ . 
\ee

\item   Also note that if  $T$ has eigenvalue $1$ and $\mu>0$  then the amplitude is infinite.
It always makes sense as a generating series, but the ``nonperturbative sum'' must be defined on 
a subdomain.

\end{enumerate}

\subsection{Some Extensions Of The Result}

\begin{enumerate}

\item \emph{Mixed Boundary Conditions}.
We can include the full set of boundary conditions by considering
the amplitude to be an element of
\be
\bar\CA \in \End( S^\bullet(\oplus_{a,b} \CO_{ab} ))
\ee
Note that $\oplus_{a,b} \CO_{ab}$ can be thought of as $\infty \times \infty$ matrices but with
a block structure so that along the diagonal we have a $1\times 1$ block, then a $2\times 2$ block
then a $3\times 3 $ block and so on so that   at the $n^{th}$ step we have an 
$\half n (n+1) \times \half n (n+1)$
square matrix with this block structure. We let $\CE_\infty =\varinjlim \CE_N$ and then 
\be
\CW = L^2(\CE_\infty)  
\ee
and   the same splitting formula above applies, now taking into account arbitrary boundary conditions.

\item  \emph{Including Closed String States}. If we allow both intervals and
circles as in- and out-going objects then we have
\be
\bar\CA \in \End( S^\bullet(\oplus_{a,b} \CO_{ab} ) \otimes S^{\bullet} \CC ) ~ , 
\ee
Including closed string states is straightforward. Each cyclic combination
of the ingoing morphisms gives a closed string state $\mu^{-1} Tr( T_c) \varepsilon_x$.
Then if we have $n_{cl}$ ingoing states $\phi_1, \dots, \phi_{n_{cl}}$ with $\phi_i = z_i \varepsilon$
we have
\be\label{eq:OpenClosed-3}
T_1 \otimes \cdots \otimes T_n \otimes \phi_1 \otimes \cdots \otimes \phi_{n_{cl}}  \mapsto
\prod_i z_i  \sum_{\sigma \in S_n}
(\prod_{c\in cycl(\sigma)} \Tr(T_c) )\mu^{-r} B_{r+n_{cl}}(\lambda) e^{\lambda}
\ee
because after the open to closed maps have been applied with effectively have $r+n_{cl}$ ingoing
closed string states which can be combined with the usual combinatorics. Recall that $r$ in this 
sum is the number of cycles of $\sigma$. 

The same manipulations as above now give the factorization of the total
open-closed amplitude with arbitrary boundary conditions:
\be
\bar\CA: S^\bullet \CC \otimes S^\bullet \CO \to S^\bullet \CC \otimes S^\bullet \CO
\ee
where $\CO = \oplus_{a,b} \CO_{ab}$, 
as
\be\label{eq:Split-Amp-OpenClosed}
\bar\CA = \Phi_{\mu} \Phi_{\mu}^\dagger
\ee
where
\be
\Phi_{\mu}: S^\bullet \CC \otimes S^\bullet \CO  \to S^\bullet \CC \otimes L^2(\CE_\infty)
\ee
is given by
\be\label{eq:Phi-ClosedOpen}
\begin{split}
\Phi_{\mu} = & \sum_{n=0}^\infty \sum_{S_{in}}  \sum_{d=0}^\infty \int_{\CE_\infty} [dH]
(d \varepsilon^\vee)^{n} \otimes_{S_{in}} e^\vee_{i_a j_a}  \\
& \sqrt{\frac{\lambda^d}{d!}} e^{-\half V_{d/\mu}(H)} \prod_{S_{in}} H_{j_a i_a} \langle H \vert \\
\end{split}
\ee
Again recall that the sesquilinear structure is given by \eqref{eq:Open-SesquiForm}  and
\be\label{eq:Closed-Sesqui}
(\varepsilon^\vee)^\dagger = \frac{\varepsilon}{\theta} \qquad \qquad (\varepsilon)^\dagger = \frac{1}{\mu} \varepsilon^\vee
\ee

As a corollary, if we consider the HH co-vector $\Psi_{HH}^\vee : S^\bullet(\CC) \otimes S^{\bullet}(\CO) \to \kappa$
evaluated on the exponentiated diagonal we get:
\be
\Psi_{HH}^\vee ( e^{\phi} \otimes e^T ) = \exp[ \frac{ \lambda e^u}{(\det(1-T))^{1/\mu}} ]
\ee
where $\phi = u \varepsilon$, thus giving an interpretation to an expression appearing in 
\cite{Gardiner:2020vjp}.

\item\emph{Closed Constrained Boundary Conditions.}
As we have mentioned above, there is a category $\fB$ of boundary conditions. 
We can consider a bordism category where the morphisms are surfaces with 
boundary with three kinds of boundaries: First, there are in- and out-going one-dimensional 
boundaries with the endpoints of the intervals connected by constrained boundaries, 
as we have thus far discussed. But now there can be closed constrained boundaries, 
labeled by objects of $\fB$, which are neither in-going nor out-going. See Figure 5 below. 

The amplitudes with closed constrained boundaries 
can be computed by the insertion of a boundary state. 
If $\beta \in Ob(\CB)$ then its associated boundary state is obtained by 
applying the open to closed map to the identity in $\CO$ to get 
(in the notation of  \cite{Moore:2006dw}): 
\be
\iota^{\beta}(1) = \frac{w_{\beta}}{\mu} \varepsilon \in \CC
\ee
where $w_{\beta} \in \IZ_+$ is the dimension of the vector bundle over the spacetime point.
Now, if we specify that the two-dimensional morphism has $L_{\beta}$ closed constrained boundary holes
of type $\beta$ then equation \eqref{eq:OpenClosed-3} gives the following component of the 
HH covector: 
\be\label{eq:OpenClosed-4}
T_1 \otimes \cdots \otimes T_n \otimes \phi_1 \otimes \cdots \otimes \phi_{n_{cl}}  \mapsto
\prod_i z_i \prod_{\beta} \left( \frac{w_{\beta}}{\mu} \right)^{L_{\beta}}  \sum_{\sigma \in S_n}
(\prod_{c\in cycl(\sigma)} \Tr(T_c) )\mu^{-r} B_{r+n_{cl}+L_\beta}(\lambda) e^{\lambda}
\ee
\noindent\rule[0.5ex]{\linewidth}{1pt}
\usetikzlibrary{decorations.markings}
\begin{tikzpicture}
   [decoration={markings, 
    mark= at position 0.5cm with {\arrow{stealth}}}
]
   \coordinate (A1) at (0,0);
   \coordinate (A2) at (0,1);
   \coordinate (A3) at (0,2);
   \coordinate (A4) at (0,3);
   \coordinate (B1) at (2,0);
   \coordinate (B2) at (2,1);
   \coordinate (B3) at (2,2);
   \coordinate (B4) at (2,3);
   \draw (A1) to [bend right=90] (A2);
   \draw (A2) to [bend right=90] (A1);
   \draw (A1) to (B1);
   \draw (A2) to (B2);
   \draw [postaction={decorate}] (A3) -- (A4);
   \draw [dash pattern=on 1pt off 2\pgflinewidth] (A3) -- (B3);
   \draw [dash pattern=on 1pt off 2\pgflinewidth] (A4) -- (B4);
   \draw [dash pattern=on 1pt off 2\pgflinewidth] (B3) to [bend right=20] (B4);
   \draw (B2) to [bend right=30] (B3);
   \coordinate (C1) at (6,0);
   \coordinate (C2) at (6,1);
   \coordinate (C3) at (6,2);
   \coordinate (C4) at (6,3);
   \draw (C4) to [bend right=20] (B4);
   \draw (B1) to [bend right=20] (C1);
   \coordinate (D1) at (8,0);
   \coordinate (D2) at (8,1);
   \coordinate (D3) at (8,2);
   \coordinate (D4) at (8,3);
   \draw (C1) to (D1);
   \draw (C2) to (D2);
   \draw [dash pattern=on 1pt off 2\pgflinewidth] (C3) -- (D3);
   \draw [dash pattern=on 1pt off 2\pgflinewidth] (C4) -- (D4);
   \draw (D1) to [bend right=90] (D2);
   \draw (D2) to [bend right=90] (D1);
   \draw (C3) to [bend right=30] (C2);
   \draw [postaction={decorate}] (D3) -- (D4);
   \draw [dash pattern=on 1pt off 2\pgflinewidth] (C4) to [bend right=20] (C3);
   \coordinate (E1) at (3,1.5);
   \coordinate (E2) at (5,1.5);
   \draw (E1) to [bend right=20] (E2);
   \coordinate (E3) at (3.3,1.4);
   \coordinate (E4) at (4.7,1.4);
   \draw (E4) to [bend right=20] (E3);
   \coordinate (F1) at (2.4,2.8);
   \coordinate (F2) at (3.4,2.8);
   \draw [green, dash pattern=on 1pt off 2\pgflinewidth] (F1) to [bend right=90] (F2);
   \draw [green, dash pattern=on 1pt off 2\pgflinewidth] (F2) to [bend right=90] (F1) node[below]{a};
   \coordinate (G1) at (4.5,2.8);
   \coordinate (G2) at (5.6,2.8);
   \draw [blue, dash pattern=on 1pt off 2\pgflinewidth] (G1) to [bend right=90] (G2);
   \draw [blue, dash pattern=on 1pt off 2\pgflinewidth] (G2) to [bend right=90] (G1) node[below]{b};
   \coordinate (H1) at (3.5,0.5);
   \coordinate (H2) at (4.5,0.5);
   \draw [red, dash pattern=on 1pt off 2\pgflinewidth] (H1) to [bend right=90] (H2);
   \draw [red, dash pattern=on 1pt off 2\pgflinewidth] (H2) to [bend right=90] (H1) node[below]{c};
\end{tikzpicture}
\paragraph{} 
\textbf{Figure 5}: \emph{An example of a 2-morphism illustrating the difference 
between constrained boundaries attached to in or out-going intervals and the closed constrained boundaries. The latter are the red, green, and blue circles.  }

\noindent\rule[0.5ex]{\linewidth}{1pt}
\item Once we allow two-dimensional morphisms with closed constrained boundaries of type $\beta$ the question arises whether we wish to sum over the insertions of closed constrained boundaries, and with what combinatorics.
One natural thing to do is sum over all numbers of constrained boundaries of all types. 

For a single fixed boundary condition $\beta$ we can consider
\be\label{eq:SumConstrain1}
\sum_{L_\beta=0}^\infty  \frac{x_\beta^{L_\beta} }{L_{\beta}!}  \bar \CA_{L_\beta} ~ . 
\ee
We introduced a fugacity $x_\beta$ so that we can recover the amplitudes with a fixed
number of constrained boundaries if desired.

The net effect is simply to replace
\be
\sqrt{\frac{\lambda^d}{d!}}  \to \sqrt{\frac{\lambda^d}{d!}} \exp[ \half \frac{x_\beta w_\beta d}{\mu} ]
\ee
in the definition of $\Phi_{\lambda}$ in equation \eqref{eq:Phi-ClosedOpen}.

Now the value of the HH covector on the exponentiated diagonal, is:
\be\label{eq:SumConstrain2}
\begin{split}
\sum_{L_\beta=0}^\infty   \frac{x_\beta^{L_\beta} }{L_{\beta}!}  
\Psi^\vee_{HH,L_\beta}(e^\phi e^T) & = \sum_{L_\beta=0}^\infty
\sum_{n_{cl}=0}^\infty \sum_{n=0}^\infty   \frac{x_\beta^{L_\beta} }{L_{\beta}!}\frac{1}{n_{cl}!} \frac{1}{n!}
  \Psi^\vee_{HH,L_\beta}(\phi^{n_{cl} }  T^n ) \\
& = \sum_{L_\beta=0}^\infty \sum_{n_{cl}=0}^\infty \sum_{n=0}^\infty
\sum_{d=0}^\infty  \frac{x_\beta^{L_\beta} }{L_{\beta}!}   \frac{1}{n_{cl}!}
\frac{\lambda^d}{d!} \left(\frac{d w_{\beta}}{\mu} \right)^{L_{\beta}}
 \left(d u\right)^{n_{cl} } \sum_{\sigma \in S_n} \left(\frac{d}{\mu}\right)^r \prod_{c} Tr T_c \\
& = \exp[ \frac{ \lambda e^u e^{x_\beta w_{\beta}/\mu}  }{(\det(1-T))^{1/\mu}} ] ~ .\\
\end{split}
\ee
When we generalize to include all boundary conditions we now
get the result
\be
 \prod_{\beta=0}^\infty \exp[ \frac{ \lambda e^u e^{x_\beta w_{\beta}/\mu}  }{(\det(1-T))^{1/\mu}} ]
\ee
%
%We now encounter (at least formally) (labeling a boundary condition by the dimension $w$ of the vector %bundle):
%
%\be
%\sum_{w=0}^\infty e^{x_w w/\mu}
%\ee
%
The boundary conditions
are labeled by vector bundles over the spacetime point, so we can  identify $\beta$ with the dimension $w_{\beta} \in \IZ_+$ of this vector space. So, now identifying a boundary condition with a nonnegative integer $w\in \IZ_+$, the grand formula for the HH covector evaluated on 
the exponentiated diagonal, including all possible closed constrained holes is: 
\be
 \exp[ \lambda e^u  \frac{ \sum_{w=0}^\infty e^{wx_w /\mu}  }{(\det(1-T))^{1/\mu}} ]
\ee
where $T$ is infinite by infinite.

\item \emph{$\dim ~ \CC >1 $}.
The decompositions
\be
\CC = \oplus_x  \IC \varepsilon_x
\ee
\be
\CO_{ab} \cong \oplus_x Mat_{N_{a_x}, N_{b_x}}
\ee
are orthogonal. So if a connected surface has ingoing homogeneous
elements with different supports $x\not=y$ then the amplitude is zero.
Therefore all the amplitudes factorize as a product over $x\in X$, 
up to the multinomial combinatorial factors described in the 1d and 
closed 2d case above.  For example, the vacuum-to-vacuum 
amplitude with $L_\beta$ constrained boundaries of type $\beta$ is given by 
\be\label{eq:ConstrainedVacVac}
\bar \CA_{\{L_\beta\}}(0,0) = 
\sum_{\vec d \in \IZ_+^n} \prod_x \left( \frac{ e^{-\lambda_x}\lambda_x^{d_x}}{d_x!} \right) \prod_{\beta} 
\left( \sum_x \frac{w_{\beta,x}}{\mu_x} d_x \right)^{L_\beta}
\ee 

\item In section \ref{subsubsec:RelationCoherent} we have suggested 
an alternative viewpoint to the ensemble viewpoint of holography. 
Nevertheless, it should be noted that, for specialized values of parameters, 
the  result \eqref{eq:ConstrainedVacVac} does raise the possibility 
of a holographic interpretation of the amplitudes in terms of 
\underline{ensembles} of 1d TQFTs, more in line with the interpretation 
of \cite{Marolf:2020xie}. The specialization means that we must take 
the values of the open string couplings $\mu_x^{-1}$ to be positive integers. 
\footnote{Of course, in this case the sum over geometries 
diverges and must be defined by analytic continuation.} 
For simplicity, assume $\CC$ is one-dimensional and we restrict 
attention to just one boundary condition with $\CO_{aa} = {\rm Mat}_{N}(\IC)$. 
Then we can write 
equation  \eqref{eq:ConstrainedVacVac} as 
\be\label{eq:EnsembleInterp}
\frac{\bar \CA_{L}(0,0)}{\bar \CA(0,0) } = \langle \CZ(S^1)^L \rangle_{\CE}
\ee 
where, on the RHS $\CZ(S^1)$ is a stochastic variable on an ensemble $\CE$
of 1d TQFT's $\CE=\{ \CT_d \}_{d\in \mathbb{N}} $ where $\CT_d(pt)$ is a 
complex vector space of dimension $d \frac{N}{\mu}$, and $\CZ(S^1)$ is the 
stochastic variable given by the partition function of the 1d TQFT 
on the circle. The measure on the ensemble is simply the Poisson 
measure $p(\CT_d) = \frac{\lambda^d}{d!} e^{-\lambda}$. It is also possible 
to view the right hand side of \eqref{eq:EnsembleInterp} as a partition 
function on the circle of a 1d TQFT whose target category is that of  
finite-dimensional vector bundles over a measure space. In order for these 
interpretations to be completely satisfactory, one would wish to extend 
them to all the other kinds of amplitudes we have discussed, in particular, 
with nonempty in- and out-going state spaces. 
We leave the exploration of this idea for another occasion.

\end{enumerate}

\subsection{What Is This Mathematical Structure? }

One may well ask whether the amplitudes we have derived
belong to a known algebraic structure. For example, it is
natural to wonder if they define interesting cyclic
cohomology classes, or $A_\infty$ or $L_\infty$ maps.
The answer to these questions is ``no.''

For the case $\dim ~\CC = 1$ and $\CO = \Mat_{N\times N}(\kappa)$ we list
the first several $n$-fold multiplcations $m_n: S^n \CO \to \CO$, which put an
interesting algebraic structure on $\CO$: 

\noindent
The $0-1$ amplitude is simply 
\begin{equation}
    m_0=\frac{I}{\mu}\frac{B_1(\lambda)}{\mu}e^{\lambda}
\end{equation}
where $I$ is the unit matrix. 
The $1-1$ amplitude is 
\begin{equation}
    m_1(T)=e^{\lambda}\bigg[\frac{I}{\mu}\frac{B_2(\lambda)}{\mu^2}\text{Tr}(T)+\frac{T}{\mu}\frac{B_1(\lambda)}{\mu}\bigg]
\end{equation}
so $m_1^2$ is nonzero if $\lambda$ is nonzero. 
The 2-to-1 map from two ingoing intervals to one outgoing interval is: 
\begin{equation}
    \begin{split}
        m_2(T_1,T_2) & = 
        e^\lambda \frac{I}{\mu}\bigg(\frac{B_3(\lambda)}{\mu^3}\text{Tr}(T_1)\text{Tr}(T_2)+\frac{B_2(\lambda)}{\mu^2}\text{Tr}(T_1T_2)\bigg)\\
        & +e^\lambda\bigg[\frac{T_1}{\mu}\frac{B_2(\lambda)}{\mu^2}\text{Tr}(T_2)+\frac{T_2}{\mu}\frac{B_2(\lambda)}{\mu^2}\text{Tr}(T_1)+\frac{T_1T_2+T_2T_1}{\mu}\frac{B_1(\lambda)}{\mu} \bigg]~ . \\
    \end{split}
\end{equation}
Note that on the right hand side of this equation $T_1T_2$ refers to the standard matrix product 
and \underline{not} an element of $S^2\CO$. Thus $T_1 T_2 + T_2 T_1$ is the standard Jordan product on $\CO$. 

The above product $m_2: S^2 \CO \to \CO$ is nonassociative. The associator 
can (after tedious computation) be shown to be 
\begin{equation}
    \begin{split}
        &a(T_1,T_2,T_3)=\frac{e^{2\lambda}}{\mu}\bigg[[[T_1,T_3],T_2]\frac{B_1(\lambda)^2}{\mu^2}+\frac{T_1}{\mu}\bigg(N\text{Tr}(T_2)\text{Tr}(T_3)\frac{B_3(\lambda)B_2(\lambda)}{\mu^5}+2\text{Tr}(T_2)\text{Tr}(T_3)\frac{B_3(\lambda)B_1(\lambda)}{\mu^4}\\
        &+N\text{Tr}(T_2T_3)\frac{B_2(\lambda)^2}{\mu^4}+4\text{Tr}(T_2T_3)\frac{B_2(\lambda)B_1(\lambda)}{\mu^3}+\text{Tr}(T_2)\text{Tr}(T_3)\frac{B_2(\lambda)^2}{\mu^4}\bigg)\\
        &-\frac{T_3}{\mu}\bigg(N\text{Tr}(T_1)\text{Tr}(T_2)\frac{B_3(\lambda)B_2(\lambda)}{\mu^5}+2\text{Tr}(T_1)\text{Tr}(T_2)\frac{B_3(\lambda)B_1(\lambda)}{\mu^4}+N\text{Tr}(T_1T_2)\frac{B_2(\lambda)^2}{\mu^4}\\
        &+4\text{Tr}(T_1T_2)\frac{B_2(\lambda)B_1(\lambda)}{\mu^3}+\text{Tr}(T_1)\text{Tr}(T_2)\frac{B_2(\lambda)^2}{\mu^4}\bigg)\\
        &+\frac{I}{\mu}(\text{Tr}(T_1)\text{Tr}(T_2T_3)-\text{Tr}(T_1T_2)\text{Tr}(T_3))\bigg(N\frac{B_2(\lambda)B_3(\lambda)}{\mu^5}+2\frac{B_1(\lambda)B_3(\lambda)}{\mu^4}+\frac{B_2(\lambda)^2}{\mu^4}\bigg)\bigg]
    \end{split}
\end{equation}

Perhaps what we are finding here is a homotopy generalization of a Jordan algebra.
We leave this question to future study. For the record we give the general 
formula for the map $m_n: S^n \CO \to \CO$:  
\begin{equation}
    \begin{split}
        &m_n(T_1,\cdots,T_n)=\sum_{d=0}^{\infty}\frac{\lambda^d}{d!}\bigg[\frac{I}{\mu}\frac{d}{\mu}A_n^{(d)}(T_1,\cdots,T_n)+\\
        &\sum_{i=1}^n\frac{T_i}{\mu}\frac{d}{\mu}A_{n-1}^{(d)}(T_1,\cdots,\hat{T}_i,\cdots,T_n)+\\
        &\sum_{i=1}^n\sum_{j>i}^N\frac{\text{Sym}(T_iT_j)}{\mu}\frac{d}{\mu}A_{n-2}^{(d)}(T_1,\cdots,\hat{T}_i,\cdots,\hat{T}_j,\cdots,T_n)+\\
        &\sum_{i<j<k}\frac{\text{Sym}(T_iT_jT_k)}{\mu}\frac{d}{\mu}A_{n-3}^{(d)}(T_1,\cdots,\hat{T}_i,\cdots,\hat{T}_j,\cdots,\hat{T}_k,\cdots,T_n)+\cdots+\\
        &\frac{\text{Sym}(T_1T_2\cdots T_n)}{\mu}\frac{d}{\mu}A_0^{(d)}\bigg]
    \end{split}
\end{equation}
where 
\be 
\text{Sym}(T_1T_2\cdots T_k) = \sum_{\sigma \in S_k} T_{\sigma(1)} \cdots T_{\sigma(k)} 
\ee
with ordinary matrix products on the right hand side and the scalar-valued 
functions $A_n^{(d)}$ are defined by 
\begin{equation}
    A_n^{(d)}(T_1,\cdots,T_n)=\sum_{\sigma\in S_n}\bigg(\frac{d}{\mu}\bigg)^{\text{cyc}(\sigma)}\sum_{i_1,\cdots,I_n}\sum_{j_1,\cdots,j_n}(T_1)_{i_1j_1}\cdots (T_n)_{i_nj_n}\delta_{i_1\sigma(j_i)}\cdots \delta_{i_n\sigma(j_n)}
\end{equation}
For example,
\begin{equation}
    \begin{split}
        &A_0^{(d)}=1 \\
        &A_1^{(d)}(T)=\frac{d}{\mu}\text{Tr}(T) \\
        &A_2^{(d)}(T_1,T_2)=\frac{d^2}{\mu^2}\text{Tr}(T_1)\text{Tr}(T_2)+\frac{d}{\mu}\text{Tr}(T_1T_2) \\
        &A_3^{(d)}(T_1,T_2,T_3)=\frac{d^3}{\mu^3}\text{Tr}(T_1)\text{Tr}(T_2)\text{Tr}(T_3)+\frac{d^2}{\mu^2}(\text{Tr}(T_1T_2)\text{Tr}(T_3)+\text{Tr}(T_2T_3)\text{Tr}(T_1)+\\
        &\text{Tr}(T_3T_1)\text{Tr}(T_2))+\frac{d}{\mu}(\text{Tr}(T_1T_2T_3)+\text{Tr}(T_1T_3T_2))
    \end{split}
\end{equation}
and so on. Finally 
the notation $A_{n-1}^{(d)}(T_1,\cdots,\hat{T}_i,\cdots,T_n)$ means that the parameter $T_i$ is absent in the string of $(n-1)$ parameters, and so on. 
\paragraph{}

\section{Higher Dimensions}\label{sec:HigherDimensions}

A natural question is whether the summed amplitude 
\eqref{eq:Sum-TQFT} makes sense in higher dimensional 
topological field theory.
 \footnote{We thank D. Freed, D. Jordan, and F. Luo for useful discussions and correspondence 
 on the issues in this section.  }
This is a very difficult 
question because above two dimensions the classification 
of manifolds becomes much more difficult. This is even 
true in three dimensions, despite the remarkable progress 
in three-manifold topology which has been achieved in 
recent decades. The essential problem, in the three-dimensional 
case, is that there is no known adequate classification of the 
fundamental groups that can appear. In the four-dimensional 
case all finitely presented fundamental groups can appear. 

It was recognized some time ago by S. Carlip \cite{Carlip:1992us,Carlip:1992wg}
that the entropy of topological complexity poses an important 
challenge to the notion that quantum gravity should 
involve a sum over topologies. 
\footnote{The sum over topologies in 3d gravity has recently been 
discussed in \cite{Afkhami-Jeddi:2020ezh,Benjamin:2021wzr,Maloney:2020nni}.}
We note here that in the 
standard examples of 3d TQFT the sum over bordisms seems 
irretrievably divergent. This is apparent already from consideration 
of the vacuum to vacuum amplitude, namely the sum over all 
bordisms from $\emptyset$ to itself. This sum has a subsum 
of 3-folds of the type $\Sigma_g \times S^1$ where $\Sigma_g$ is a compact 
oriented surface of genus $g$. But by general principles of 
TQFT we always have $\CZ(\Sigma_g \times S^1)= \dim \CZ(\Sigma_g)$. 
Thus the sum \eqref{eq:Sum-TQFT} above contains the 
sub-sum 
\be\label{eq:3d-SubSum}
\sum_{g=0}^\infty  \dim~ Z(\Sigma_g) ~ . 
\ee
In standard examples of 3d TQFT, such as Chern-Simons-Witten theory for a compact 
gauge group $Z(\Sigma_g)$ grows with $g$ (i.e. the dimension grows with topological 
complexity). One might try to resort to $\zeta$-function regularization to 
define \eqref{eq:3d-SubSum} but it is worth noting that for the important special 
case of invertible theories even then one would be computing $\zeta(1)$ which is 
irretrievably infinite. 

One might hope that the vacuum to vacuum sum is an overall divergence which somehow
can be factored out, but alas, this is not the case. In the standard examples of 
3d TQFT the state space $\CZ(S^2)$ is the one-dimensional algebra over $\kappa$. 
It follows that the 
state created by cutting out a ball from $\Sigma_g \times S^1$ is 
$\dim(\CZ(\Sigma_g))$ times the identity element of $\CZ(S^2)$. 
Thus, a sub-sum of the three-dimensional analog of the sum over ``handle-adding elements'' 
is again divergent. 

The above remarks illustrate that the generalization to higher dimensions is difficult. 
But, it might not be impossible. 
We see three logical possibilities for constructing higher-dimensional analogs 
of the sum over bordisms. We list them in order of increasing plausibility: 

\begin{enumerate}
    \item Perhaps the sum over bordisms is a conditionally convergent sum and there are massive cancellations making the full sum convergent.  Well, maybe. 
    \item Perhaps there are interesting sub-categories of the bordism category, or categories of manifolds with extra structure and accompanying TQFTs where the sum is more manageable. We are not aware of any examples. 
\footnote{One natural way to attempt a manageable sub-category would be to restrict to 3-folds which 
are handlebodies. But this fails to produce a proper sub-category since every closed 3-fold admits 
a Heegaard decomposition. The problem is that we must allow gluing, and when gluing we must 
allow arbitrary elements of the mapping class groups of surfaces. It appears to us that this is a 
fundamental difficulty with proposals in the literature to define 3d quantum gravity by restricting 
to sums over handlebodies. }
    \item One might hope that there are 3d TQFT's where the partition functions and state spaces \underline{vanish} at large topological complexity. This hope is dashed by the following elementary observation of Sergei Gukov: If the statespace 
    of a 3d TQFT vanishes on a Riemann surface of \underline{any} genus, then all the amplitudes of the theory are zero. 
    
\end{enumerate}

\section{Future directions}

It would be interesting to verify the splitting phenomenon we have observed in 
other versions of open-closed 2d TQFT. In this paper we explored some non-semisimple 
TQFTs but did not give a completely general discussion. 
Moreover, a TQFT with a global symmetry can be 
 coupled to background, nondynamical, $G$-gauge fields using the description of such theories in terms of Turaev algebras, as spelled out in \cite{Moore:2006dw}. It is asserted (without detailed proof)  in \cite{Moore:2006dw} that, after summing over bundles the amplitudes will be those of an ordinary 2d TQFT - the orbifold theory. The sum over bordisms can be organized 
 as a sum over bundles for fixed topology of the worldvolume, and then a sum over the topologies. Therefore the $G$-equivariant 
 theories will also admit a splitting. It is possible that the $G$-equivariant theories without a sum over bundles provide an interesting explicit example of extended defects and it would be interesting to explore to what extent one has a splitting 
 formula when including such defects. 
 \footnote{We thank Shu-Heng Shao for raising this question.}
Some of the results of 
 \cite{deMelloKoch:2021lqp} might be helpful in exploring this question. 
 Similarly, summing over bordisms for 2d spin TQFT's has been considered in \cite{Balasubramanian:2020jhl}, but the total amplitudes were not fully worked out. Finally, it would be preferable to give a general conceptual explanation of the existence of a splitting formula, rather than merely verifying it case-by-case in a number of examples, as we have done here.  
 
%
%We have seen that the total amplitudes can be expressed in terms of a splitting formula, $\bar{\mathcal{A}}=\Phi\Phi^{\dagger}$. %But  $\Phi:S^{\bullet}\mathcal{C} \to\mathcal{W}$
%is not unique, since it can be right multiplied by a unitary. 
%More importantly, the space $\CW$ is not unique, 
%and this raises the question of whether one can define a
%\underline{minimal} splitting space.    
%

It is natural to ask if the kind of splitting formula we have observed for 1d and 2d TQFT generalizes to 2d topological string theory, or even for perturbative string theory proper. Similarly, one could ask if it extends to JT gravity. 
We have not explored these extensions. One extension we have explored is the case of coupling the TQFT to 2d Yang-Mills theory. 
It appears that the splitting property does hold in this case, under suitable conditions, as we will explain in a separate 
publication. 

Finally, as discussed above, there are some possible routes to extending these ideas to higher-dimensional TQFT's. While this appears challenging, a successful outcome could be of great interest.

\appendix

\section{Bell Polynomials}\label{App:BellPolynomials}

The exponential Bell polynomial
$B_n(x_1,x_2,\dots, x_n)$ counts the number of ways a set of $n$ elements can be partitioned
into parts. The coefficient of $x_1^{k_1} \cdots x_n^{k_n}$ counts the number of ways we have
$k_1$ subsets of cardinality 1, $k_2$ subsets of cardinality $2$, ...  $k_j$ subsets of cardinality $j$.

We have the useful formula
\be\label{eq:Bell-Def1}
B_n(x_1,x_2,\dots, x_n) = \left( \frac{\partial}{\partial t} \right)^n \exp[ \sum_{j=1}^\infty x_j \frac{t^j}{j!} ] \vert_{t=0}
\ee
So
\be
\begin{split}
B_1 & = x_1 \\
B_2 & = x_1^2 + x_2 \\
B_3 & = x_1^3 + 3 x_1 x_2 + x_3 \\
B_4 & = x_1^4 + 6 x_1^2 x_2 + 3 x_2^2 + 4 x_1 x_3 + x_4 \\
\end{split}
\ee

Note that $B_n(x)$, defined by putting $x_1 = x_2 = \cdots = x_n = x$ satisfies
\be
\begin{split}
e^x B_n(x) & = e^x  \left( \frac{\partial}{\partial t} \right)^n \exp[ x e^t - x  ] \vert_{t=0}\\
 & =  \left( \frac{\partial}{\partial t} \right)^n  \sum_{d=0}^\infty \frac{(x e^t)^d}{d!}  \vert_{t=0}\\
 & = \sum_{d=0}^\infty \frac{x^d}{d!} d^n\\
 \end{split}
\ee

Notably, the Bell polynomials $B_n(x)$ occur naturally in
the theory of the harmonic oscillator as expectations of the number operator in a
coherent state. Let $\Psi_x : =  \exp[ x a^\dagger] \vert 0 \rangle$ where we
normalize $[a,a^\dagger] = 1$.
Then
\be\label{eq:HO-vev}
\langle \Psi_x , N^n \Psi_x \rangle = e^{\vert x \vert^2 } B_n( \vert x \vert^2)
\ee

\section{Action of $b_\vee^m$ On Symmetric Products}\label{App:Apply-b}

 When applying powers of these operators to symmetric products the following technical
consideration is important to bear in mind.   On the tensor algebra it makes sense to define
\be
b^\vee:  T^\bullet V \rightarrow V^\vee \otimes T^\bullet V
\ee
by acting on the first factor:
\be
b^\vee ( \phi_1 \otimes \cdots \otimes \phi_n) := b^\vee(\phi_1) \otimes \phi_2 \otimes \cdots \otimes \phi_n
\ee
That's the result of attaching a cylinder bent into a U-shape to convert an outgoing circle to an
ingoing circle.

Our amplitudes are much more naturally written in the symmetric algebra and we can
define $b^\vee: S^\bullet V \to V^\vee \otimes S^\bullet V$ by requiring commutativity
of the diagram:
\be
{\xymatrix{ T^\bullet V \ar[r]^{b^\vee} \ar[d]_{\pi}  &  V^\vee \otimes T^\bullet V \ar[d]^{1\otimes \pi} \\
S^\bullet V \ar[r]^{b^\vee}  & V^\vee \otimes S^{\bullet} V \\ }
}
\ee
Thus, when we wish to evaluate $b^\vee(\phi_1 \cdots \phi_n)$ where $\phi_1 \cdots \phi_n \in S^nV$ we
take the canonical lift to $T^{n} V$:
\be
\frac{1}{n!} \sum_{\sigma \in S_n} \phi_{\sigma(1)} \otimes \cdots \otimes \phi_{\sigma(n)}
\ee
and apply $b^\vee$ there to get
\be
\frac{1}{n!} \sum_{\sigma \in S_n} b^\vee( \phi_{\sigma(1)})  \otimes \cdots \otimes \phi_{\sigma(n)}
\ee
Now split the sum on $\sigma$ into $n$ terms where $\sigma(1) = j$, $j=1, \dots, n$. This partitions
$S_n$ into $n$ torsors for $S_{n-1}$. Projecting down to $V^\vee \otimes S^{n-1} V$ we learn
that if $\phi_1 \cdots \phi_n \in S^n V$ then
\be
b^\vee(\phi_1 \cdots \phi_n) = \frac{1}{n} \sum_{j=1}^n b^\vee(\phi_j) \otimes \left(
\phi_1 \cdots \widehat{\phi_j} \cdots \phi_n\right)
\ee
where $\left(\phi_1 \cdots \widehat{\phi_j} \cdots \phi_n\right) \in S^{n-1} V$ is gotten by omitting the factor $\widehat{\phi_j}$.

 Now we can continue applying $b^\vee$ so that
\be
b^\vee \otimes \cdots \otimes b^\vee:  S^\bullet V \to ( V^\vee)^{\otimes m} \otimes S^\bullet V
\ee
but we are interested in the projection to $S^m(V^\vee) \otimes S^{\bullet} V$. We get the formula:
\be\label{eq:bvee-rule}
(b^\vee)^m (\phi_1 \cdots \phi_n) = {n \choose m}^{-1} \sum_{S_1 \amalg S_2} \left( \prod_{i\in S_1} b^\vee(\phi_i)  \right)
\otimes \left( \prod_{j\in S_2} \phi_j \right)
\ee
where the sum is over all distinct disjoint decompositions $S_1 \amalg S_2 = \{1, \dots, n\}$ where $\vert S_1 \vert = m$
and these are \underline{unordered} sets.

\section{Simple Remark On Multi-Linear Functionals}\label{App:MultiLinear-Diagonal}

  In general, a \underline{multi-linear and totally symmetric} function on $F: S^n V\to \kappa$ for
any vector space $V$ is completely determined by its values on the diagonal,
$G(v):=F(v,\dots, v)$. We can recover $F(v_1, \dots, v_n)$ from the formula
\be
F(v_1, \dots, v_n) = \frac{1}{n!}  \frac{\p}{\p t_1}\vert_0  \cdots \frac{\p}{\p t_n}\vert_0 G(t_1 v_1 + \cdots + t_n v_n)
\ee
Now, if we have a series of multi-linear and totally symmetric functions $F_n: S^n V \to \kappa$ we
can assemble them into one single function $F = \oplus_n F_n : S^\bullet V \to \kappa$. Then we define
\be
G(v):= F(e^v) := \sum_{n\geq 0} \frac{1}{n!} F_n(v, \dots, v)
\ee
and we recover
\be
F_n(v_1, \dots, v_n) = \frac{\p}{\p t_1}\vert_0  \cdots \frac{\p}{\p t_n}\vert_0 G(t_1 v_1 + \cdots + t_n v_n)
\ee
(Note there is no $1/n!$ in this formula.)

\end{document}